# Investigate the impact of PTMs on the protein backbone conformation.

Pierrick Craveur[1,2,3,4,5,#], Tarun J. Narwani[1,2,3,4,#], Joseph Rebehmed[1,2,3,4,6,+]
& Alexandre G. de Brevern [1,2,3,4,+,*]

[1] INSERM, U 1134, DSIMB, F-75739 Paris, France.
[2] Univ Paris, Univ de la Réunion, Univ des Antilles, UMR_S 1134, F-75739 Paris, France.
[3] Institut National de la Transfusion Sanguine (INTS), F-75739 Paris, France.
[4] Laboratoire d'Excellence GR-Ex, F-75739 Paris, France.
[5] Department of Integrative Structural and Computational Biology, The Scripps Research Institute, La Jolla, California, USA.
[6] Department of Computer Science and Mathematics, Lebanese American University, Byblos 1h401 2010, Lebanon.

* Corresponding author: Dr. de Alexandre G. de Brevern, INSERM UMR_S 1134, DSIMB, Université de Paris, Institut National de Transfusion Sanguine (INTS), 6, rue Alexandre Cabanel, 75739 Paris cedex 15, France
e-mail : alexandre.debrevern@univ-paris-diderot.fr
Tel: +33(1) 44 49 30 38 / Fax: +33(1) 47 34 74 31

Key words: rigidity, mobility, deformability, N-glycosylation, phosphorylation, methylation, statistics, renin endopeptidase, liver carboxylesterase, cyclin-dependent kinase 2 (CDK2), actin.

# Abstract

Post-Translational Modifications (PTMs) are known to play a critical role in the regulation of the protein functions. Their impact on protein structures, and their link to disorder regions have already been spotted on the past decade. Nonetheless, the high diversity of PTMs types, and the multiple schemes of protein modifications (multiple PTMs, of different types, at different time, etc) make difficult the direct confrontation of PTM annotations and protein structures data.

We so analyzed the impact of the residue modifications on the protein structures at local level. Thanks to a dedicated structure database, namely PTM-SD, a large screen of PTMs have been done and analyze at a local protein conformation levels using the structural alphabet Protein Blocks (PBs). We investigated the relation between PTMs and the backbone conformation of modified residues, of their local environment, and at the level of the complete protein structure. The two main PTM types (N-glycosylation and phosphorylation) have been studied in non-redundant datasets, and then, 4 different proteins were focused, covering 3 types of PTMs: N-glycosylation in renin endopeptidase and liver carboxylesterase, phosphorylation in cyclin-dependent kinase 2 (CDK2), and methylation in actin. We observed that PTMs could either stabilize or destabilize the backbone structure, at a local and global scale, and that these effects depend on the PTM types.

# Introduction

After its synthesis, a protein can undergo reversible or irreversible covalent modifications, namely Post-Translational Modifications (PTMs). In some cases, such as N-glycosylation, the modifications take place during the translation, but are commonly included under the term PTMs. The modifications alter the physicochemical properties of the proteins and thereby regulate enzymatic activity, cellular localization and intermolecular interactions (Deribe et al. 2010; Duan and Walther 2015; Zhao et al. 2010). This wide range of functions is reflected by the high diversity of PTMs depending on the cell type, the tissue, and the organism in which proteins are synthesized. Additionally a protein could be modified in many ways and at different residue positions over time. The same position may also undergo changes of different kinds. However, changes may be specific to certain amino acids, like N-glycosylation found on asparagine in the specific consensus sequence Asn-X-Ser/Thr where X can be any amino acid residue but proline (Moremen et al. 2012). PTMs are extremely diverse, ranging from the addition of a small group of atoms, such as phosphorylation (Humphrey et al. 2015), to the attachment of bulkier oligosaccharide by glycosylation (Imberty 1997). PTMs are essential to regulate biological functions, such as DNA transcription by histone methylation and demethylation, acetylation or phosphorylation (Bannister and Kouzarides 2011; Mijakovic et al. 2016), nuclear-cytosolic or extra-cytosolic transport by SUMOylation (Hendriks and Vertegaal 2016; McIntyre et al. 2015) or glycosylation (Dewald et al. 2016; Imberty and Perez 1995), tagging proteins for degradation by ubiquitination (Zhou and Zeng 2016), and regulation of kinase activity with phosphorylation (Krupa et al. 2004). PTMs are also associated with major human diseases such as cancer, diabetes, cardiovascular disorders and Alzheimer's disease (Kamath et al. 2011; Li et al. 2010; Martin et al. 2011).

In context of protein function, this diversity leads to cooperative mechanisms of PTMs such as competition for serine and threonine residues between phosphorylation and O-glycosylation (Butt et al. 2012; Zeidan and Hart 2010); ubiquitination favored over phosphorylation leading to protein degradation (Vodermaier 2004), or the interactions between PTMs regulating the activity of the p53 protein and Histones (Brooks and Gu 2003; Latham and Dent 2007). These observations suggest towards the existence of a PTM-code (Creixell and Linding 2012; Minguez and Bork 2017; Nussinov et al. 2012), which is based on the presence and association of several PTMs leading to the realization of particular functions. Recently, the increasing number of annotations on PTMs have assisted scientists to study the cross talk or direct / indirect influences among different types of PTMs (Lu et al. 2011; Tokmakov et al. 2012; van Noort et al. 2012) their competition for the same residue (Danielsen et al. 2011), or the co-evolution of different PTMs sites within the same protein (Minguez and Bork 2017; Minguez et al. 2013; Minguez et al. 2012).

Many databases and prediction tools have been developed to enhance the understanding of various PTMs in different organisms and to simplify the analysis of complex PTM data (Gianazza et al. 2016). These PTM databases contain crucial sequence annotations, specific to some PTM types and/or organisms (Gupta et al. 1999; Hornbeck et al. 2015; Yao and Xu 2017), and provide related structural data which mainly correspond to the mapping of the PTM sites in protein entries of the Protein Data Bank (PDB) (Huang et al. 2016). Numerous machine learning methods consisting of predicting PTM sites were published recently. They mainly differ in the types of PTM and/or organisms focused, in their learning protocols (support vector machine, random forest, neuronal network, etc.), and in the set of descriptors extracted from the mining of the experimental data (Audagnotto and Dal Peraro 2017; Gianazza et al. 2016). Few of them, used descriptors derived from structural data, such as prediction of secondary structures, disorder and accessible surface area (Lopez et al. 2017; Lorenzo et al. 2015), or from structural properties extracted from PDB (Torres et al. 2016;

Wuyun et al. 2016).

The proteins functions and their 3D structures are intrinsically related. Hence, it is expected that PTMs, which regulate function, impact the structure of proteins as well. Several previous studies have investigated the effects that PTMs could have on the protein structure and dynamics, using X-ray data (Xin and Radivojac 2012), and NMR data (Gao and Xu 2012). Xin and Radivojac (Xin and Radivojac 2012) computed local and global RMSDs between modified (with at least one PTM), and unmodified PDB chains of the same protein. They concluded from the statistical analysis of their RMSDs that N-glycosylation and phosphorylation induce conformational changes, with a limited impact, at both local and at global levels, with a larger influence for phosphorylation. On their side, Gao and Xu (Gao and Xu 2012) suggest that disorder-to-order transition could be induced by the modifications of phospho-serine/-threonine, various types of methyllysines, sulfotyrosine, 4-carboxyglutamate, and potentially 4-hydroxyproline.

Disorder regions are mainly defined as series of missing residues in X-ray structures taken from the PDB, they are highly frequent (more than 80% of the X-ray structures with a resolution worse than 1.75 Å have missing residues) (Djinovic-Carugo and Carugo 2015). Many analyses have focused on Intrinsic Disorder Proteins (IDPs, (Piovesan et al. 2017)) that are often implicated in molecular recognition. They bind a partner molecule and undergo "induced folding" or "disorder-to-order transition" leading to an ordered state in which PTMs can paly a role (Fuxreiter and Tompa 2012). Hence IDP regions have been associated with numerous PTMs, as hydroxylation, methylation, and notably phosphorylation (Gao and Xu 2012; Vucetic et al. 2007; Xie et al. 2007a; Xie et al. 2007b) which was recently proposed to function as protein interaction switches in more ordered regions (Betts et al. 2017).

To investigate a potential relation between the presence of a PTM on the protein and the structure itself, we used PTM-SD (Craveur et al. 2014) that gives access to X-ray structures of modified residues in proteins that specifically correspond to all PTM annotations. We investigated the impact of PTMs on the protein backbone conformations observed in crystallographic data. First, the diversity of the backbone conformations of N-glycosylated and phosphorylated regions was analyzed. Then local and global effects in the backbones were compared between 4 specific examples of PTMs associated to a high number of experimental data. Finally, the presence and absence of PTMs on the protein were also compared in regards to the backbone flexibility.

## Methods

**PTM-SD**. Post Translational Modification Structural Database (http://www.dsimb.inserm.fr/dsimb_tools/PTM-SD/) is designed to give users a curated access to the proteins for which one or more Post Translational Modification(s) is (are) structurally resolved in the Protein Data Bank (PDB) and also experimentally annotated in dbPTM (Huang et al. 2016) and PTMCuration (Khoury et al. 2011). PTM-SD uses diverse set of rules to underline the discrepancies between annotation in the structure and the sequences owing to different sources. Also, PTM-SD allows the user to create customized PTMs queries and perform different analyses on the returned entries, *e.g.* computing distribution of organisms, proteins, PDB codes/chains, and PTM types, assigning PBs, computing $N_{eq}$ (see bellow), highlighting discrepancies between PDB sequence and UniProt sequence, clustering for generation of non-redundant dataset, etc.

Besides a global view on PTMs, the database also provides details for each PTM and further connects to different PTM information and annotations found in other databases. Such data are very informative for studying relationship between PTMs and protein structures, for designing comparative modeling protocol, and for prediction protocol based on different approaches, for example, on secondary structure descriptors.

**Dataset.** The dataset used in this study is generated using PTM-SD. It comprises of structures pertaining to phosphorylation, N-glycosylation and Methylation while also contains corresponding structures without a modification (a tabular summary of the dataset can be found in Table S1 and the dataset in Dataset S1). The comprehensive dataset included a total of 9.870 PTMs present on 5.948 structures. From these, 7.110 modifications were N-glycosylation while 1.874 were phosphorylation and 886 methylations. The dataset was further refined to remove redundancy (>25% identity) using PTM-SD clustering toolkit. It is important to generate non-redundant dataset by filtering over-represented chains to avoid biasing the analyses..

The non-redundant dataset consisted of 348 N-glycosylation on 156 PDB chains from 41 different organisms, 92 phosphorylations on 76 structures from 12 different organisms and 19 methylations on 15 structures from 9 distinct organisms. Similar datasets were generated, using PTM-SD, for the analysis of different types of phosphorylations. 57 serine modifications on 45 pdb chains while 29 phosphothreonine and 34 phosphotyrosine are found on 29 and 30 unique pdb chains (tabular details are provided in Table S2).

A derived dataset was also created to assess the impact of PTM on the global structure. Therefore, a dataset comprising 4 proteins with the largest number of observations; Renin endopeptidase (N-glycosylation), Liver carboxylesterase (N-glycosylation), Cyclin dependent Kinase 2 (Phosphothreonine) and Actin (Methylation) was generated (refer to Table S3), sequences with too many missing residues were not taken into account. As we are working with same protein chain, it is essential to keep here all entries, i.e. to have a better view of the (potential) local protein conformation differences.

**Protein Blocks.** Protein Blocks (PBs) are a structural alphabet composed by a set of 16 local prototypes of 5 residues in length, labeled from *a* to *p* (see Figure 1A). They are described as series of eight Φ, Ψ dihedral angles. An unsupervised classifier similar to Kohonen Maps (Kohonen 1982; Kohonen 2001) and Hidden Markov Models (Rabiner 1989) obtained them. Briefly described, PBs *m* and *d* are prototypes for the central region of α-helix and β-strand, respectively. PBs *a* to *c* primarily represent the N-cap of β-strand while *e* and *f* correspond to C-caps; PBs *g* to *j* are specific to coils, PBs *k* and *l* correspond to N cap of α-helix while C-caps are represented by PBs *n* through *p*.

The Protein Blocks efficiently approximate all local regions of a protein structure with an average RMSD of 0.41 Å (Etchebest et al. 2005). They have been employed in various approaches including protein superimposition (Gelly and de Brevern 2011; Joseph et al. 2012), structural analysis (Dudev and Lim 2007; Wu et al. 2010) or prediction (Rangwala et al. 2009; Zimmermann and Hansmann 2008) of protein binding sites, and structural analysis of β-bulges (Craveur et al. 2013).

***Protein Blocks assignment.*** The assignment translates a 3D structure to 1D sequence of PBs. In our study input structures come from PDB files. The algorithm uses 5 residues long window for each position. For each "$n^{th}$" position, 8 dihedrals $\psi_{n-2}$, $\varphi_{n-1}$, $\psi_{n-1}$, $\varphi_n$, $\psi_n$, $\varphi_{n+1}$, $\psi_{n+1}$, $\varphi_{n+2}$ are compared to the reference set of 16 PBs. The comparison is performed using the RMSDA criteria (*Root Mean Square Deviation on Angular values*) (Schuchhardt et al. 1996):

$$RMSDA(V_1, V_2) = \sqrt{\frac{\sum_{i=1}^{i=M-1}\left[\psi(V_1) - \psi(V_2)\right]^2 + \left[\varphi(V_1) - \varphi(V_2)\right]^2}{2(M-1)}} \qquad (1)$$

$V_1$ is the 8 dihedrals vector extracted from the $M = 5$ residues long window; $V_2$ is the 8 dihedrals vector corresponding to the compared PBs. PB, which gets lowest RMSDA is chosen as the representing conformation observed in the window. PB assignment was done with a modified version of the PBxplore tool (https://github.com/pierrepo/PBxplore, (Barnoud et al. 2017))(see Figure 1B).

**$N_{eq}$ - *Local structure entropy.*** 3D structures of a specific protein could be observed with different conformations in X-ray crystals, or during molecular dynamics simulations. This could be attributed to the intrinsic flexibility of the structure or the consequences of interactions with small molecules (ligand, cofactor, water molecules), or macromolecules (proteins, DNA, RNA). Under such scenarios, each of these 3D conformations would be assigned a different PB sequence (see Figure 1B). By analyzing the variation of PBs at each position, it's possible to investigate the local conformational changes in a protein structure.

The equivalent number of PBs ($N_{eq}$) is a statistical measurement similar to Shannon entropy and represents the average number of PBs observed at a given position (de Brevern et al. 2000). $N_{eq}$ is calculated as follows:

$$Neq = exp(-\sum_{x=1}^{16} f_x \ln f_x)\ \ \ \ \ \ \ \ \ \ (2)$$

where $f_x$ is the frequency of PB $x$ ($x$ takes values from $a$ to $p$). A $N_{eq}$ value of 1 indicates that only one type of PB is observed, while a value of 16 is equivalent to a random distribution.

For example $N_{eq}$ value around 6 would indicate that at the current position of interest, 6 different PBs are observed. If $N_{eq}$ exactly equal to 6, this means that 6 different PBs are observed in equal proportions (1/6). By plotting the computed $N_{eq}$ value at each residue position (see Figure 1B), it is possible to locate which protein regions have local conformation change, or in other words, which region of the structure represents backbone deformation. A clear interest of $N_{eq}$ in regards to the use of root mean square deviation (see (Burra et al. 2009) for similar analyses) is to provide a simple measure that quantifies locally the divergence.

**B-factor normalization.** B-factor values are partly dependent on the resolution of the crystal and of the refinement process (Hinsen 2008; Linding et al. 2003; Schlessinger and Rost 2005). Also crystallographic contact packing and addition of stabilizing molecules can impact the B-factor values. Then, in order to compare B-factor from several X-rays of the same protein, it is needed to normalize the value. In our study raw B-factor values were normalized as recommended by Smith et al (Smith et al. 2003), starting by removed outliers values detected with the median-based approach. The normalized B-factors are computed as follow:

$$Bfactor_i^{norm} = \frac{Bfactor_i^{raw} - \mu}{\sigma} \quad (3)$$

where $\mu$ and $\sigma$ are the mean and the standard deviation of the B-factor values (without outliers) respectively, $Bfactor_i^{norm}$ is the normalized B-factor at position $i$ in the sequence (and the structure) and $Bfactor_i^{raw}$ the original B-factor value.

**Various analyses.** The 3D structure representations were generated using PyMOL software (http://www.pymol.org) [The PyMOL Molecular Graphics System, Version 1.7 Schrödinger, LLC.] (Delano 2013). Most of the analyses were done using Python programming language and R software (R Core Team 2013).

# Results

**Backbone protein conformational diversity at the vicinity of N-glycosylated and phosphorylated residues.** Using PTM-SD (Craveur et al. 2014), the two most frequent PTMs were focused upon, N-glycosylation and phosphorylation. 3,092 and 1,307 chains were found containing 7,110 N-glycosylations and 1,873 phosphorylations in 100 and 22 organisms respectively. A non-redundant dataset, with less than 25% of identity between the corresponding UniProt sequences, was generated, resulting in the selection of 348 N-glycosylations (for 156 protein chains in 41 organisms) and 92 phosphorylations (for 75 protein chains in 12 organisms, see Supplementary Table S1).

$N_{eq}$ was used to analyze the local protein conformations. Based on 16 PBs, it underlines the diversity of local conformation in a finer manner than the classical secondary structures (see Methods). Figure 2 shows the variations of PBs around the two PTMs. N-glycosylated and phosphorylated sites do not exhibit any significant preferences for a particular local structure conformation. The $N_{eq}$ values are very high, ranging from 9.03 to 11.44 for N-glycosylation, and from 5.95 to 11.41 for phosphorylation, implying that these two modifications are observed in widely diverse structural contexts. Nonetheless, it is interesting to note that both types of PTMs have an overall different $N_{eq}$ profiles (see black curve in Figure 2).

For N-glycosylation, the PTM site position presents an $N_{eq} = 10.76$ which is extremely high. This would mean that N-glycosylated residues have backbone conformation as diverse as $^2/_3$ of the backbone conformations observed in proteins. Additionally the surrounding positions of the PTM sites show the same level of diversity, with $N_{eq}$ values fluctuating around 10.

For phosphorylation, the $N_{eq}$ profile is quite different. First of all, as indicated by the red curve on Figure 2, the surrounding positions of phosphorylation sites are mainly disordered. The farther the positions are from the PTM sites, the higher is the disorder in the structure; meaning that less residues were available at these positions in the PDB chains to be used for the PBs assignments and the $N_{eq}$ computation. However the data used is diverse enough to reach high level of $N_{eq}$ (6.48) computed at the PTM position. Preceding positions -8 to -2 show even higher diversity. It is important to confirm that the absence of data in the surrounding positions is not the consequences of phosphorylation sites located at the N- or C-terminus; indeed only 12 of them (out of 92) are close to the protein extremities.

A more precise analysis of the distribution of each type of PBs (see Supplementary Figure S1) underlines that N-glycosylation and phosphorylation sites are observed for all types of local conformations, almost any kind of PBs (except PBs *g for both,* and, *h, j,* and *p* for phosphorylation).

The conformations of the N-glycosylation sites and their surrounding residues are mainly associated with the PBs *d* and *m*. However, this proportion does not exceed 31%. It is interesting to note that the positions +3 to +6 following the N-glycosylation sites are significantly observed in a PB *d* conformation. This illustrates the fact that in ~1/3 of the times N-glycosylation site precedes a β-strand conformation.

For phosphorylation, the modification sites have a preference of PB *d*, the cores of β strands, in a little over 40% of cases. The vicinity of the phosphorylation sites is also observed with a wide variety of conformations, however a slight preference was observed for the PBs *b, c, d, f, l* and *m*. It should be noted that more than 50% of the phosphorylation sites are separated by two residues of a PB *d*.

It is important to understand that data used here provides information on the backbone conformation of PTM sites when the modifications are present, but do not obviously reflects the backbone in the absence of modifications.

Additionally, while phospho-serine and phospho-threonine share similar PB profiles, they are distinct from phospho-tyrosine (see Supplementary Figures S2, S3, and Supplementary Table S2). The modified residues were observed in a large set of backbone conformations for all three cases, but the preferences for the core β-strand conformation (PB *d*) is greater in the case of Serine and Threonine.

.

***Local backbone diversity compared to global backbone diversity in modified proteins.*** In order to compare the flexibility of the PTM region with the rest of the protein, we selected a large number of 3D chains corresponding to the same protein. Each chain was solved with a single PTM at identical sequence positions. 4 different proteins were studied, covering 3 types of PTMs: N-glycosylation in renin endopeptidase and liver carboxylesterase, phosphorylation in cyclin-dependent kinase 2 (CDK2), and methylation in actin. A total of 471 PDB chains were used in this analysis (see Supplementary Table S3).

The $N_{eq}$ profiles of modified sites and surrounding positions were compared with those of all other positions in the proteins. Figure 3 shows the example of one N-glycosylated residue, at position 141, in renin endopeptidase. Figure 3A is a zoom around the PTM site, while Figure 3B shows the $N_{eq}$ all along the protein. In this example, the maximum entropy is found at position 234, with a $N_{eq}$ value of 7.13. This position and its surroundings are associated with the maximum number of missing residues (red curves in Figure 3B), underlining a highly flexible region. It corresponds to what Zhang et al. defined as a Dual Personality Fragments (DPF): a protein region, which appears either ordered or disordered in crystal structures. It is suggested that DPFs are potential targets of regulation by allostery or PTMs (Zhang et al. 2007). Here, this flexible region (position 230-238, see red fragment in Figure 3C) is not annotated as PTM site, and also does not interact with ligands in the structures; but interestingly it includes 4 positions (230 to 234) known to be missing in a second isoform of this protein.

In comparison, the backbone of the modified residue is always ordered and presents slight deformations with $N_{eq}$ of 1.94. Its immediate neighbor positions are in the same range, with slightly higher values in positions -6, -1, and +1 ($N_{eq}$ values 2.58, 2.10, and 2.53 respectively).

In Figure 3A, the PTM site seems to be slightly more deformable than majority of its surrounding positions. Using a larger scale (Figure 3B) this deformation does not seems to be significantly different than other deformable parts along the sequence. To quantify it precisely, statistical tests were performed for each case (see Table 1, Figure 4, Supplementary Figures S4 and S5).

Firstly, the Shapiro-Wilk (SK) test provides extremely low *p*-value, in all the cases, forcing the rejection of the null hypothesis (see columns 3 and 4 of Table 1). This underlines that $N_{eq}$ values for PTM-region and the rest of the protein does not follow a normal distribution. Further, the nonparametric Mann-Whitney-Wilcoxon test was used to see if $N_{eq}$ profiles observed in the PTM-region are significantly different from those observed in the rest of the protein. With a type I error α = 5%, only the phosphorylated Thr-160 in the Cyclin-dependent kinase 2 protein and its neighboring positions have a significantly different $N_{eq}$ profile than the rest of the protein; the *p*-value being equal to 0.012.

It should be noted that in both cases of N-glycosylation, no significant differences were observed between the $N_{eq}$ profile of the PTM-region and the $N_{eq}$ profile of the rest of the protein.

In comparison to $N_{eq}$, the B-factor does not give a measure of the deformation of the backbone, but could be used to represent its mobility in the crystal context. For each of the 4 proteins of interest, the B-factors of the Cα were extracted from every PDB chain. After normalization (see Methods), the B-factors were averaged for each, structurally available, position along the sequence. The same statistical analyses, as applied to $N_{eq}$, were performed with the B-factors in order to compare the backbone mobility in the PTM areas, and in the rest of the protein (see Supplementary Figure S6 and Supplementary Table S4).

***Local and global backbone diversity compared between modified and unmodified***

***proteins.*** To determine if $N_{eq}$ profiles observed on all positions are due to the presence of the PTM in the structure, the $N_{eq}$ calculations were also performed on X-ray structures without PTMs (see Supplementary Table S3).

For two of these proteins, the renin endopeptidase (P00797) and the carboxylesterase (P23141), the local and global $N_{eq}$ profiles, observed in presence of the N-glycosylation, are very similar to the one computed in their absence (see Figure 3, Supplementary Figures S4, S7 and S8). This highlights that the attached glycans on the structures do not impact the intrinsic flexibility of these two proteins.

This is supported by the Mann-Whitney-Wilcoxon test (see Table 1), which do not provides very low p-values when comparing local and global $N_{eq}$ profiles. An explanation could be the presence of several disulphide bonds stabilizing the structures (indicated in blue sphere in Figure 3C and S4C). Also, glycans are solvent oriented that limits their interactions with the rest of the protein.

On the contrary, the other two proteins, CDK 2 (P24941) and actin (P68135), present $N_{eq}$ profiles significantly different when the residues, Thr 160 and His 75 respectively, are not modified (see Figure 4 and Supplementary Figures S5, S9 and S10). This is also supported by the Mann-Whitney-Wilcoxon test (see Table 1).

For actin, the PTM of interest is the methylation of the Histidine 75 (H75). In this case, the both profiles with and without the PTM (see Supplementary Figures S5 and S9 respectively) show peaks at similar positions. But these peaks differ in intensity in a way that the whole actin protein seems to be more rigid when the PTM is present (lower $N_{eq}$ values). However, the majority of the actin structures used for the analysis interact with ligands. These ligands are from different types (such as ADP, ATP, swinholide A, pectenotoxin 2, etc, see Supplementary Figure S11), and bind at multiple sites. The lack of consistency in the number, the types and the binding modes of the ligands across the structures make it difficult to conclude on the exclusive effect of H75 methylation. Nonetheless, from a local point of view, the ligands do not interact with the modified Histidine and its direct surroundings. At the -1 and +1 positions (see Supplementary Figures S5A and S9A), a clear difference in the $N_{eq}$ value could be observed. The values increase from 1.96 to 3.95 for position -1, and from 1.63 to 2.43 for position +1. Regardless the overall rigidity of the backbone, the methylation increases the deformation in the direct neighborhood of the PTM site.

The analysis of cyclin-dependent kinase 2 structures indicates an effect of the presence of the PTM on the flexibility. Here the PTM of interest is the phosphorylation of threonine 160. This modification is always observed in X-ray structures when the kinase proteins are complexed with protein partners. Therefore, to compare $N_{eq}$ profiles with PTM and without PTMs only chains complexed with Cyclin A2 have been used (see Table 2). As for actin, most of the kinase X-ray structures used (except 3) interact with ligands (see Supplementary Figure S12). However in this case, the ligands bind in the same pocket, regardless that the structures are phosphorylated or not. The consistent binding location makes it possible to compare $N_{eq}$ profiles, in order to investigate the PTM effect in flexibility.

Comparison of both $N_{eq}$ profiles (with and without modification) shows some significant differences. A striking one is observed at the site of modification (position 160) and its surrounding positions. In the absence of the phosphorylation (see Supplementary Figure S10), this area corresponds to a wide flexible loop, sometimes not even ordered in few structures (see red line). Nonetheless, in the presence of the phosphorylation, the $N_{eq}$ value in position +1 drops from 5.27 to 2.40, indicating an increase in rigidity. On the contrary, the region between positions 8 and 18 shows an increase of flexibility when the modification is present; values change from 3.89 to 6.52 for the two highest points. These points correspond to the direct vicinity of two other phosphorylation sites (annotated) at Thr-14 and Tyr-15.

From a functional point of view, the phosphorylation of Thr-160 is known to promote kinase activity, whereas the phosphorylation of Thr-14 and Tyr-15 are known to reduce its activity (Gu et al. 1992; Welburn et al. 2007). Thus the change in flexibility observed in these three phosphorylation sites, affirms the hypothesis of a regulatory mechanism in CDK2 involving coupled stiffening - flexibility pathway, an indicative of underlying allostery. A second area centered on the Thr-39 shows the exact same behavior as the first region. In the same way the phosphorylation of Thr-160 lead to an allosteric effect, which increases the flexibility of this region. The position 39 is also a phosphorylation site, which is not related to the kinase activity, but is implicated in the CDK2 intracellular localization and the cell apoptosis (Maddika et al. 2008). It is also interesting to note that Tyr-19, also annotated as phosphorylation site (Oppermann et al. 2009), does not undergo significant change in flexibility; the $N_{eq}$ value goes from 1.0 to 1.06 respectively in the absence and presence of the PTM.

Analyses of normalized B-factors (see Supplementary Figures S6 and S13) for all cases do not show any significant deviation. The presence of any of the three PTM types studied here (N-glycosylation, methylation, and phosphorylation) does not affect the mobility of the whole structure in crystals. The Mann-Whitney-Wilcoxon test supported it (see Supplementary Table S4). However in a local point of view, we can observe a variation of normalized B-factor from 1 to 2 on the surrounding of the phosphorylation site in CDK 2 (see Supplementary Figures S6C and S13C). It is interesting to note that, in this position, the decrease in mobility (B-factor) is associated with a decrease in deformability, namely $N_{eq}$.

## Discussion

As noted in different seminal papers (Berezovsky et al. 2017), the variety of known PTMs are around half-thousand, even if everybody agrees on their importance, questions about precise occurrence, impact on biological functions, implication in the evolution, and even modifying enzymes for many PTMs are yet to be answered (Sirota et al. 2015).

Using various datasets extracted from PTM-SD (Craveur et al. 2014), we were able to analyze the effects of some specific PTMs on protein backbone conformation. We first measured the conformational backbone diversity of modified residues and its close neighborhood positions for the two most frequent PTMs, N-glycosylation and phosphorylation. Secondly, we focused our analyses on 4 different proteins, and compared the local and global backbone diversity observed in X-ray data when one PTM is present. Finally, for these 4 cases, the backbone diversity with and without the PTM was compared.

In a non-redundant dataset of X-ray structures, N-glycosylation sites do not present particular structural characteristics related to the presence of the modification. In other words, the asparagine residues do not adopt a different backbone conformation when the glycans are linked in crystals. However, it is interesting to note that in ~1/3 of the times the consensus sequence Asn-X-Ser/Thr just precedes a local β-strand conformation.

In the case of phosphorylation, the modified residues were also observed in a wide set of backbone conformations, but exhibits preference for the β-strand conformation (PB *d*) in the case of Serine and Threonine. On the contrary, Phospho-Tyrosine doesn't exhibit any preferences.

For the 4 examples of protein with single PTM (N-glycosylation in renin endopeptidase and liver carboxylesterase, phosphorylation in CDK 2, and methylation in actin), we observed that only the phosphorylation site and its neighborhood positions display a backbone diversity that is significantly different than the rest of the proteins. By comparing with structures of unmodified CDK2, the presences of the phosphorylation on the activation loop at Thr160 have several local effects. It rigidified (potential stabilization) the backbone (lower $N_{eq}$ and lower B-factor) locally while increasing the deformation (potential destabilization) of two other regions, near Thr14-Tyr15, and near Thr39, three other phosphorylation sites related to CDK2 activity and subcellular localization respectively. The observed rigidity in the backbone is in agreement with the proposition made by Xin and Radivojac (Xin and Radivojac 2012) that phosphorylation, by introducing new H-bond and salt bridges in the local neighborhood leads to a conformational shift of the lowest valley in the energy landscape of the protein. This decrease of energy was also observed by Groban and coworkers (Groban et al. 2006) in their attempt to predict the conformation changes of the CDK2 activation loop induced by the phosphorylation using computational methods. This methodology was quite inventive, using a hierarchical loop prediction algorithm optimizing dihedral angle backbone sampling, linked to rotamer-based side chain optimization, and an all-atom force field and a Generalized Born solvation model to predict the structural consequences of phosphorylation. Finally Gao and Xu (Gao and Xu 2012) suggest, by analyzing NMR structures, that disorder-to-order transition might be introduced by Threonine phosphorylation.

In our datasets numerous PDB structures lack coordinates for some regions, which was depicted by a dip of the red curves in $N_{eq}$ plots. These particular regions mainly correspond to disorder regions in protein, which diversify the functional spectrum of proteins (DeForte and Uversky 2016) by expanding their partnered interactions with other proteins, collectively termed as protein-protein interactions (PPI). The selectivity of interacting partners and order-disorder transition of the protein structures is regulated by PTMs, and most of the times by phosphorylations. (Hsu et al. 2013). During our analysis of CDK2, we also found identical structures with missing coordinates in the catalytic loop (functional domain), flagged as a Dual Personality Fragment (Zhang et al. 2007). However, this region gets ordered based on the phosphorylation of threonine 160. We may suggest that the number of phosphorylations, in and around the catalytic domain, may also impact this selectivity of interacting partners for

CDK2.

The example of actin methylation shows an opposite effect of the modification compared to the example of CDK2. The methylation induces a local increase of the backbone diversity at the PTM site region, highlighting a higher deformation of this part of the protein. Interestingly no effect on the intrinsic mobility of this region has been observed (same B-factor profiles with or without the PTM). Unfortunately, the large variability of ligands found in the X-ray data of actin used in this study, does not allow suggesting an effect of the methylation at a global scale.

Finally for the two examples of N-glycosylation, we concluded that the addition of the glycan neither impact the local, neither global, backbone conformation of the proteins.

Despite the intrinsic linked between PTM and protein function, the molecular effects of the modifications on the protein structures and dynamics remains poorly understood. Our study, like previous systematic studies of structural data of modified and unmodified protein (Gao and Xu 2012; Xin and Radivojac 2012), shows that these effects could be of multiple types (stabilization and destabilization), at different scales (at the local PTM region, in other part of the protein as allosteric effect, or at a global level), and depend of the PTM types. However, in order to propose general rules for the molecular impact of each type of PTMs, additional structural data related to the large amount of PTM annotations already available are needed. In the scope of a systematic study, these data have to be used carefully. Indeed, many factors, independent of the presence of PTMs, could have affected the structure of the proteins, such as the crystallographic packing, the presence of engineer mutations or cross-links to help crystallization process, the presence of ions, ligands and protein partners in contact with the protein structure of interest. It must be noticed that a recent paper provides interesting insight in how to look at B-factors that could be very useful for future studies (Carugo 2018).

Molecular modeling of PTMs combines with molecular dynamic simulation is an interesting alternative. Some recent computational studies have investigate the effect of PTMs (Audagnotto and Dal Peraro 2017) on the stability of specific proteins, but the growing success of these kind of simulations also rely upon the growing number of experimental data, for the development of accurate PTM force field parameters.

## Acknowledgments

This work was supported by grants from the Ministry of Research (France), University of Paris Diderot, Sorbonne Paris Cité, National Institute for Blood Transfusion (INTS, France), Institute for Health and Medical Research (INSERM, France). PC acknowledges grant from French Ministry of Research. Calculations were done on SGI cluster granted by *Conseil Régional - Ile de France* and INTS (SESAME Grant). The authors were granted access to high performance computing (HPC) resources at the French National Computing Center CINES under grant no. c2013037147 funded by the GENCI (Grand Equipement National de Calcul Intensif). TN and AdB acknowledges to Indo-French Centre for the Promotion of Advanced Research / CEFIPRA for collaborative grant (number 5302-2). This study was supported by grant from Laboratory of Excellence GR-Ex, reference ANR-11-LABX-0051. The labex GR-Ex is funded by the program *Investissements d'avenir* of the French National Research Agency, reference ANR-11-IDEX-0005-02. This work is supported by a grant from the French National Research Agency (ANR): NaturaDyRe (ANR-2010-CD2I-014-04) to JR and AdB.

The funding bodies have no role in the design of the study and collection, analysis, and interpretation of data and in writing the manuscript.

## Ethical statements

All authors have been personally and actively involved in substantive work leading to the manuscript, and will hold themselves jointly and individually for its contents.

This material has not been published in whole or in part elsewhere.

The manuscript is not currently being considered for publication in another journal.

**LEGENDS:**

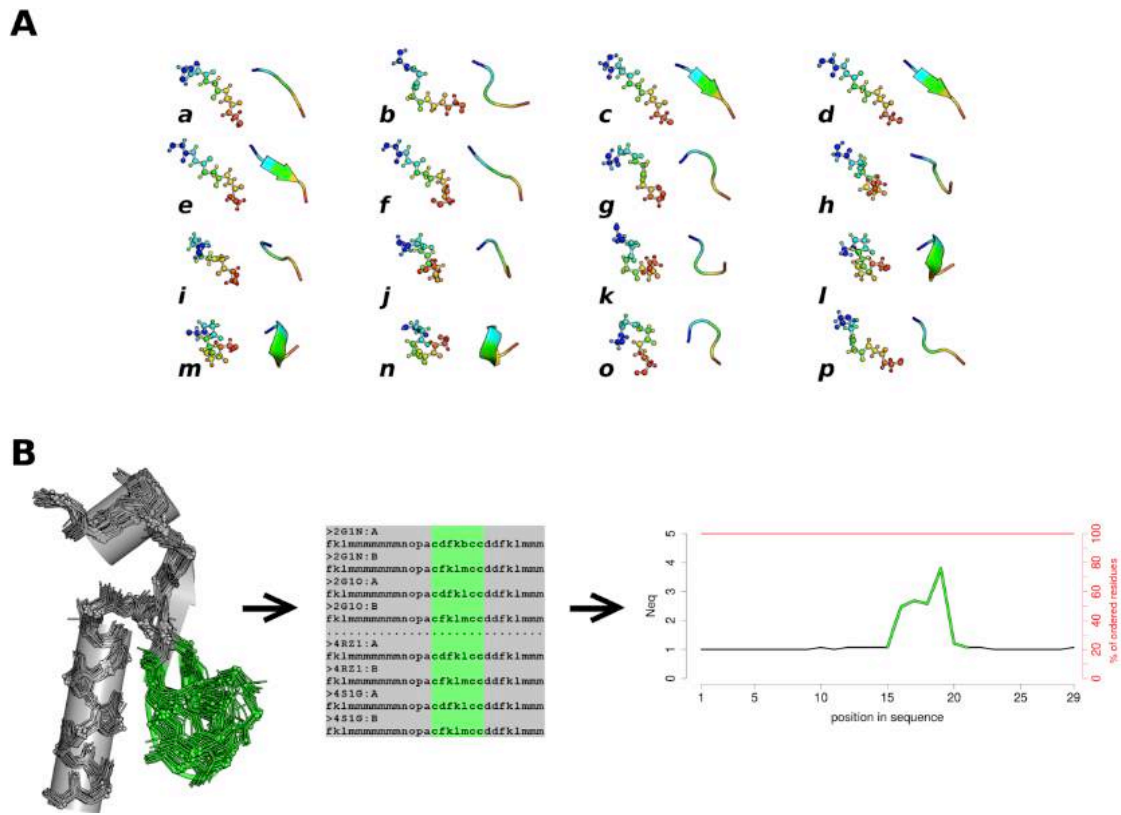

**Figure 1.** *The Protein Blocks (PB) structural alphabet and the Neq.* (**A**) The structural alphabet is composed of 16 PBs, labeled from *a* to *p*. Each PB represents the backbone conformation of a fragment of 5 residues in length, here showed in ball-and-stick and cartoon representations and colored from blue to red, from N-ter to C-ter respectively. (**B**) $N_{eq}$ is an entropic measure of the backbone conformation, taking value from 1 to 16 (see Methods). Here the backbone conformations of the deformable loop (in green) could be described with numerous sequences of PBs. Plotting the $N_{eq}$ on a graph provides an easy way to compare the backbone diversity of a protein along the sequence.

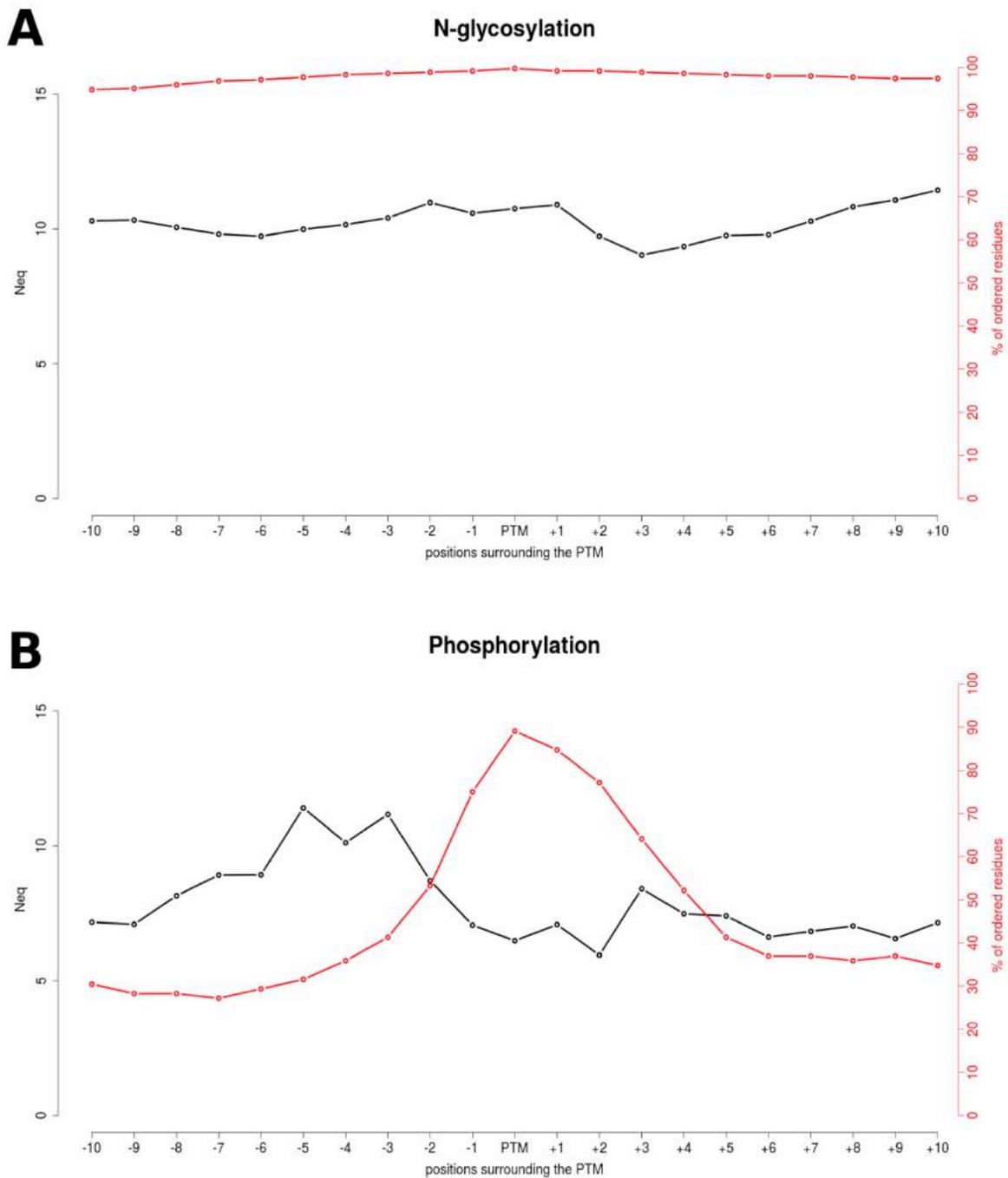

**Figure 2**. *Neq profile for N-glycosylation and phosphorylation sites*. The $N_{eq}$ profile is shown in black in the vicinity of N-glycosylation (**A**) and phosphorylation sites (**B**). The red lines indicate the amount of data used to compute $N_{eq}$ values, or in other words the percentage of ordered residues at each position in the X-ray crystals.

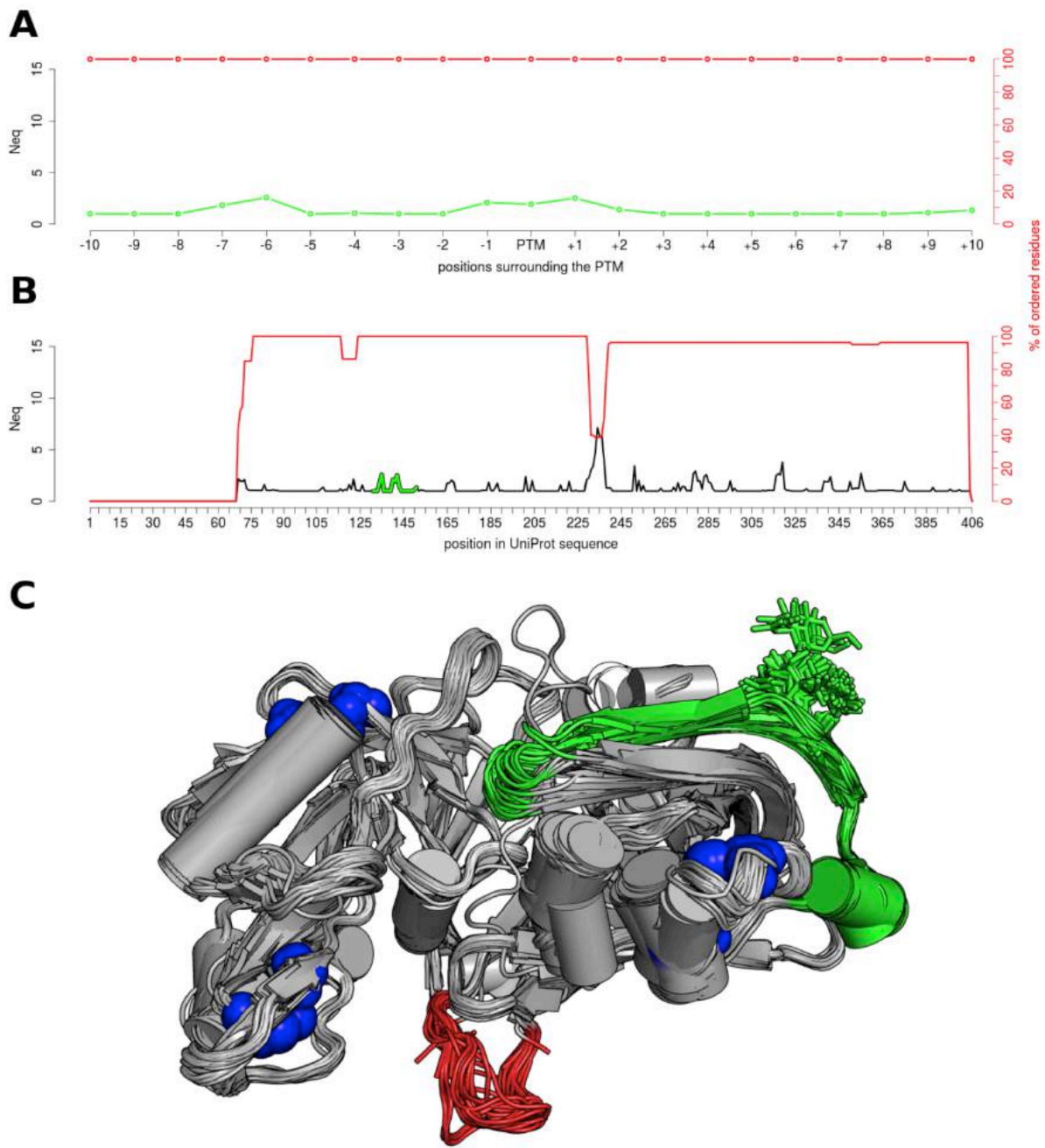

**Figure 3**. *N-glycosylation on the Asn141 of the human renin endopeptidase (P00797)* The $N_{eq}$ profiles are given at a local scale (**A**), for the surrounding positions of the PTM site (color in green), and at a global scale (**B**), for all sequence positions. (**C**) The 80 structures used for the computation were aligned on the backbone, and represented in cartoon. The glycosylated position is shown in green sticks, the disulphide-bridges in blue spheres, and the DPF loop color in red.

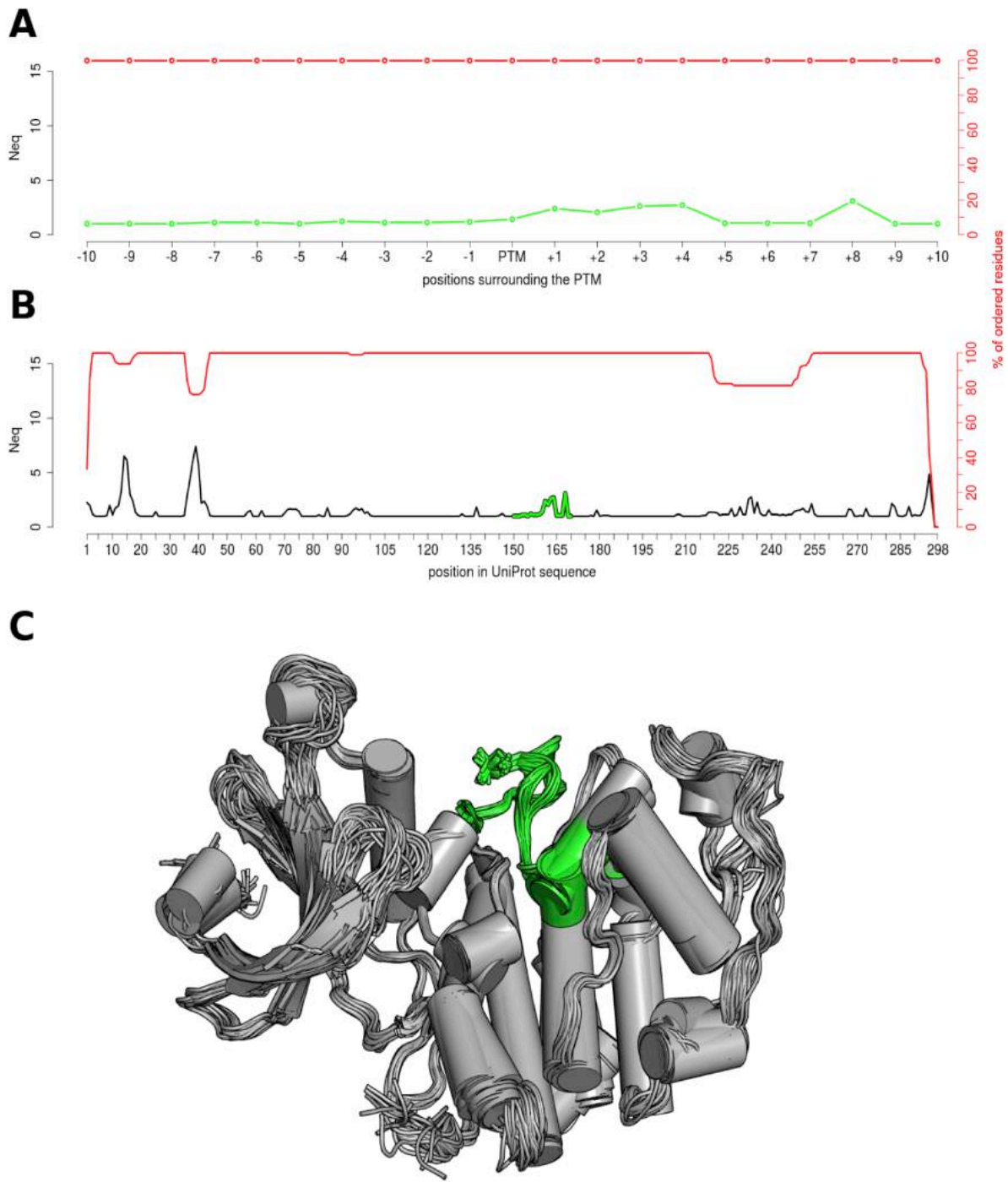

**Figure 4**. *Phosphorylation on Thr160 of human cyclin-dependent kinase 2 (P24941)* The $N_{eq}$ profiles are given at a local scale (**A**), for the surrounding positions of the PTM site (coloured in green), and at a global scale (**B**), for all sequence positions. (**C**) The 96 structures used for the computation were aligned on the backbone, and represented as cartoon. The phosphorylated position is shown in green sticks.

**Table 1**. *Comparison of structures with or without PTMs*. Statistical tests for the 4 proteins. Shapiro-Wilk (SW) test checks if data follows Normal law distribution, while Mann-Whitney-Wilcoxon (MWW) is a nonparametric test that compared mean values. Are indicated the size of samples (*n*), the calculated statistics (stats), and the p-values. The chosen risk α is equal to 5%, the significant p-values allow dismissing the hypothesis $H_0$ and are in italics.

| | | SW *Neq* PTM region vs Normal law | SW *Neq* rest of the protein vs Normal law | MWW *Neq* PTM region vs *Neq* rest of the protein | SW *Neq* all protein with 0 PTM vs Normal law | SW *Neq* all protein with 1 PTM vs Normal law | MWW *Neq* 0 PTM vs *Neq* 1 PTM |
|---|---|---|---|---|---|---|---|
| Renin endopeptidase P00797 (Human) N-glycosylation | *n* | 21 | 316 | 21/316 | 364 | 337 | 364/337 |
| | Statistic | 0.6790 | 0.4312 | 3571 | 0.5276 | 0.4429 | 59812.5 |
| | *p*-value | *1.52E-05* | *2.16E-30* | 5.08E-01 | *4.52E-30* | *5.34E-31* | 5.13E-01 |
| Liver carboxylesterase 1 P23141 (Human) N-glycosylation | *n* | 21 | 507 | 21/507 | 529 | 528 | 529/528 |
| | Statistic | 0.5522 | 0.4383 | 5582 | 0.4762 | 0.4436 | 138025.5 |
| | *p*-value | *6.40E-07* | *6.52E-37* | 6.19E-01 | *1.26E-36* | *2.12E-37* | 6.60E-01 |
| Cyclin-dependent kinase P24941 (Human) phosphorylation | *n* | 21 | 275 | 21/275 | 296 | 296 | 296/296 |
| | Statistic | 0.6792 | 0.4102 | 3687 | 0.4927 | 0.4299 | 39350 |
| | *p*-value | *1.53E-05* | *5.10E-29* | *1.94E-02* | *3.06E-28* | *1.44E-29* | *1.32E-02* |
| Actin P68135 (Rabbit) methylation | *n* | 21 | 350 | 21/350 | 371 | 371 | 371/371 |
| | Statistic | 0.7082 | 0.7201 | 3758 | 0.6745 | 0.7174 | 78367.5 |
| | *p*-value | *3.49E-05* | *8.32E-24* | 8.53E-01 | *4.34E-26* | *1.42E-24* | *7.17E-04* |

**Table 2.** *Occurrence of the different chains of human Cyclin-dependent Kinase 2 (Id: P24941) in different cyclin A2 complexes.*

| UniProt AC of the protein partner in complexes | Protein partner's name and organisms | Number of kinase chains involved |
|---|---|---|
| Complexes **with** phosphorylation on Thr 160 | | |
| P20248 | Cyclin-A2 Homo sapiens (Human) | 70 |
| P30274 | Cyclin-A2 Bos taurus (Bovine) | 24 |
| P51943 | Cyclin-A2 Mus musculus (Mouse) | 4 |
| | | Total: 98 |
| Complexes **without** phosphorylation on Thr 160 | | |
| P20248 | Cyclin-A2 Homo sapiens (Human) | 46 |
| | | Total: 46 |

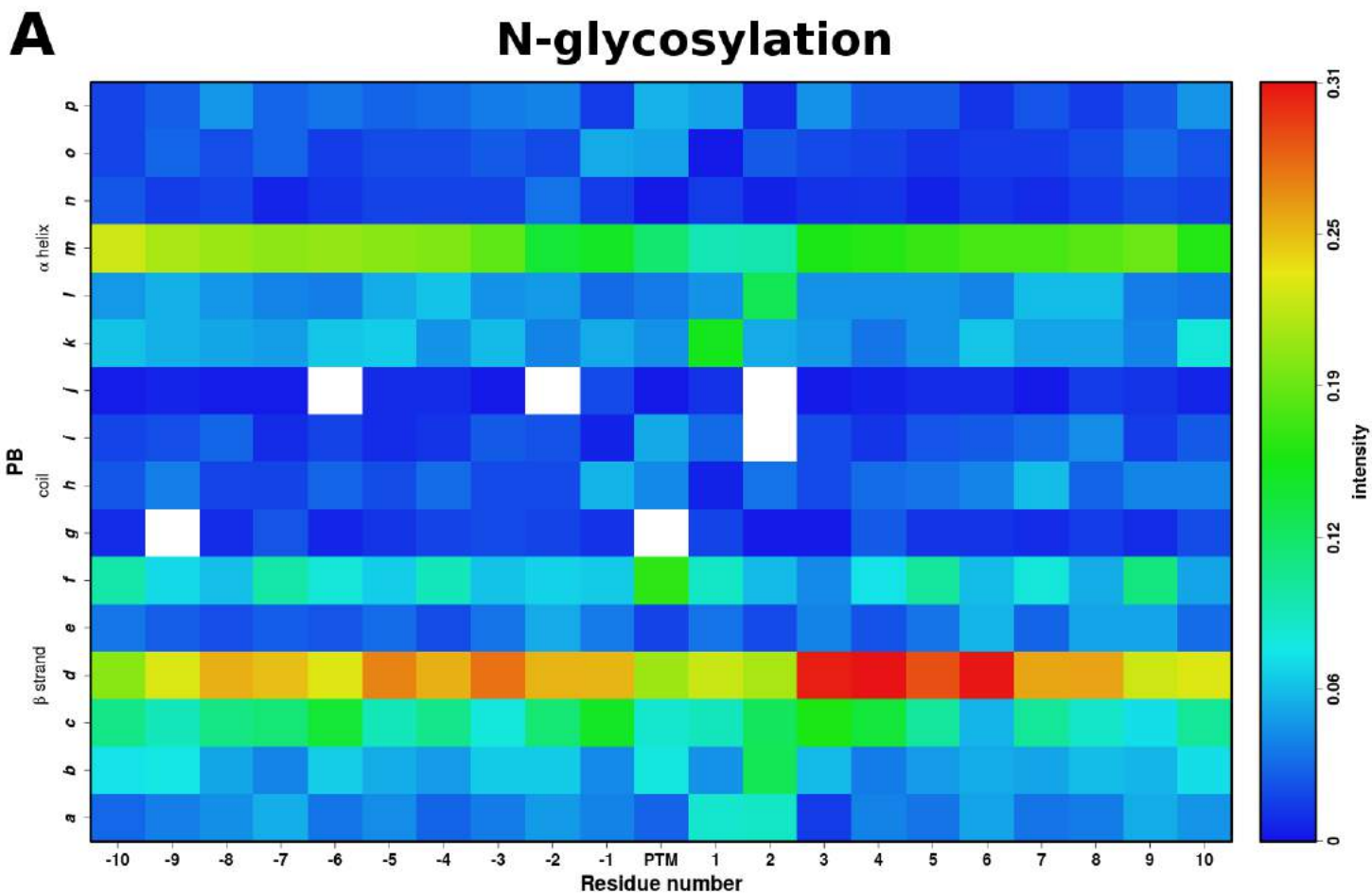

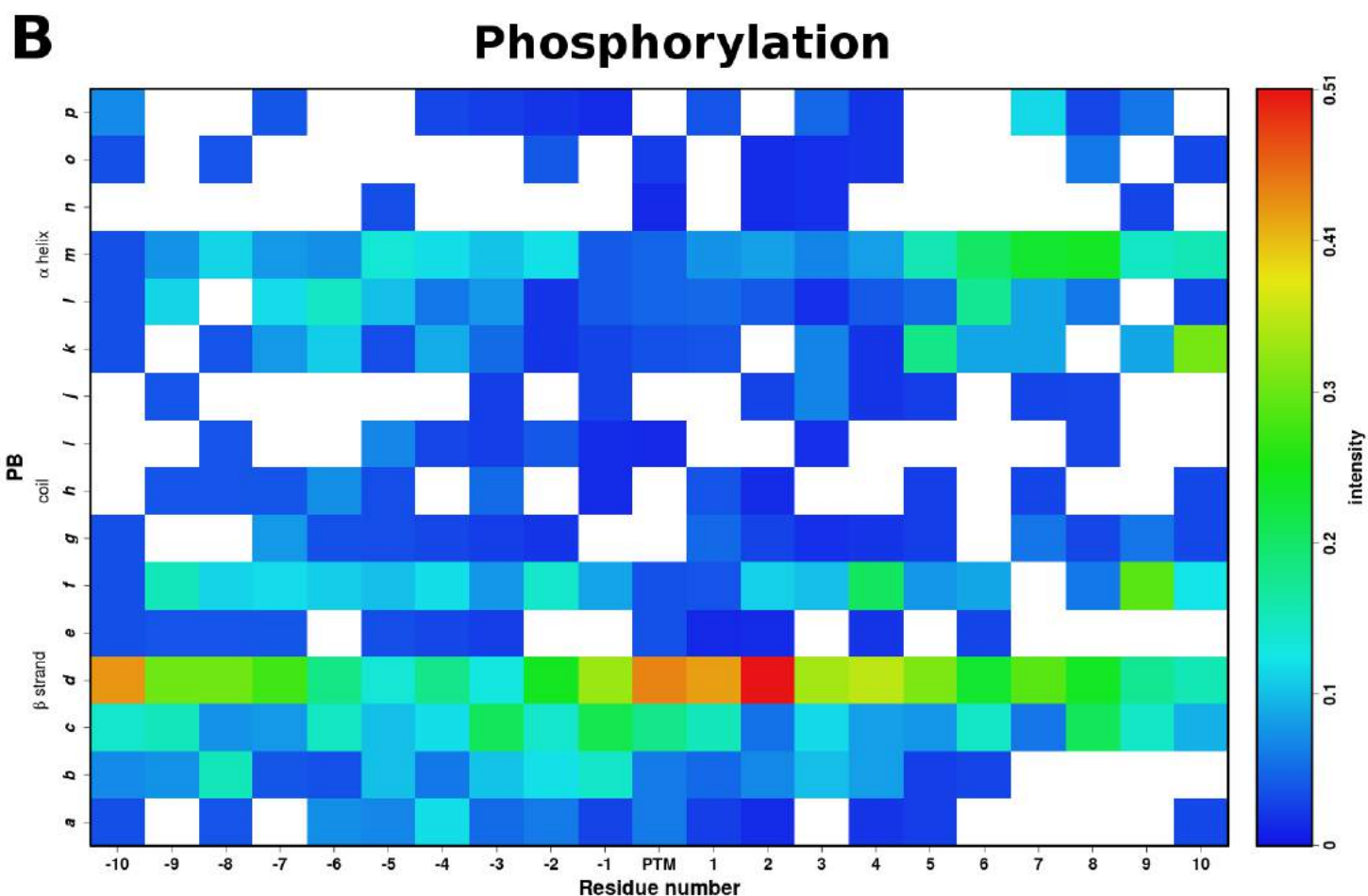

**Figure S1**. *Protein Blocks profile*. The PB distributions of the complete analysis in case of **A)** N-glycosylation and **B)** Phosphorylation. The color is represented as per the intensity legend provided on the right side. Red represents very high presence of the PB while blue being the opposite.

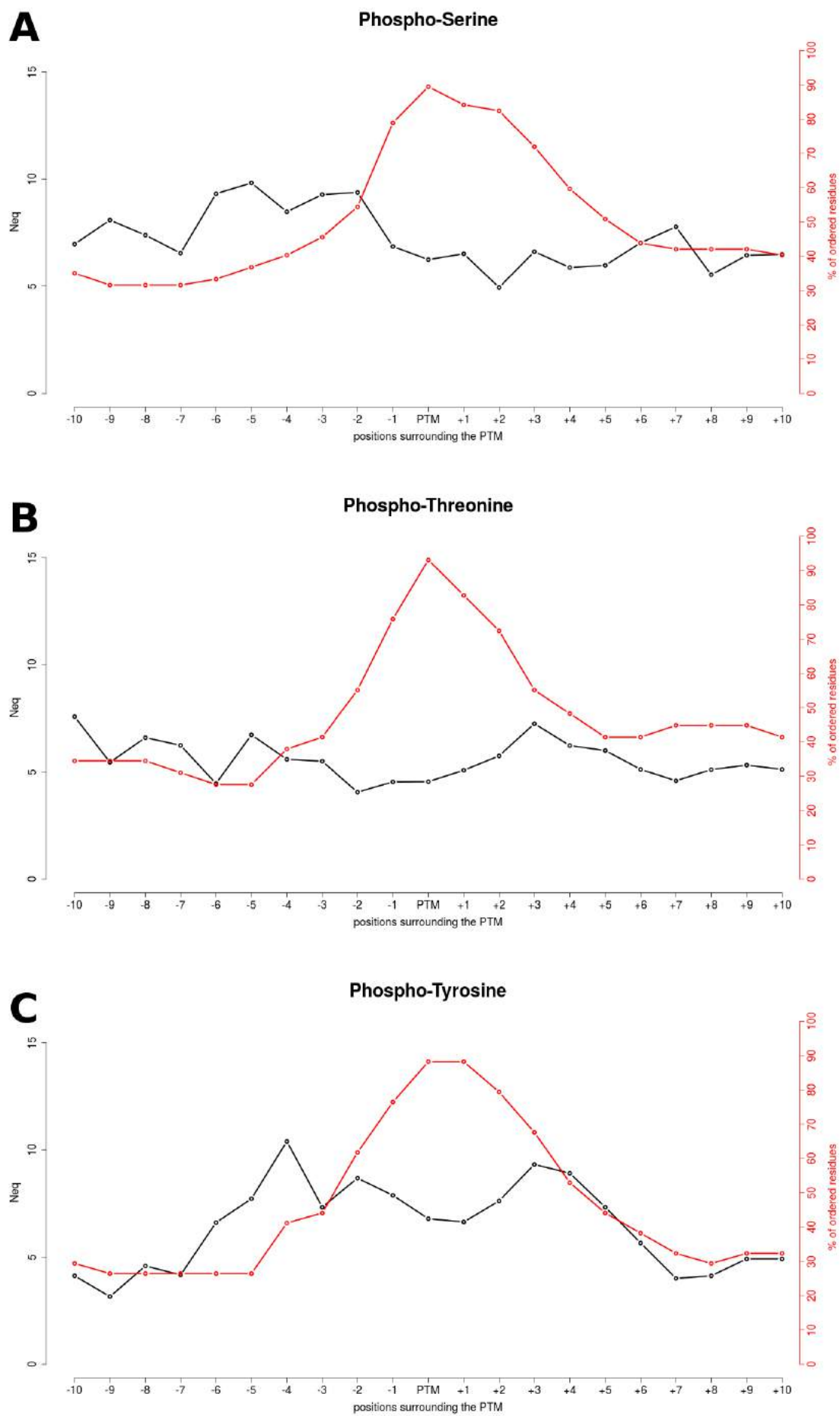

**Figure S2**. *Flexibility profile for different Phosphorylations.* $N_{eq}$ distribution curves (black) gives an insight into the extent of local structural changes at the phosphorylation site and its sequential neighbourhood. The red curve represents the percentage of available data for calculating $N_{eq}$ at a position. Higher the percentage better is the confidence.

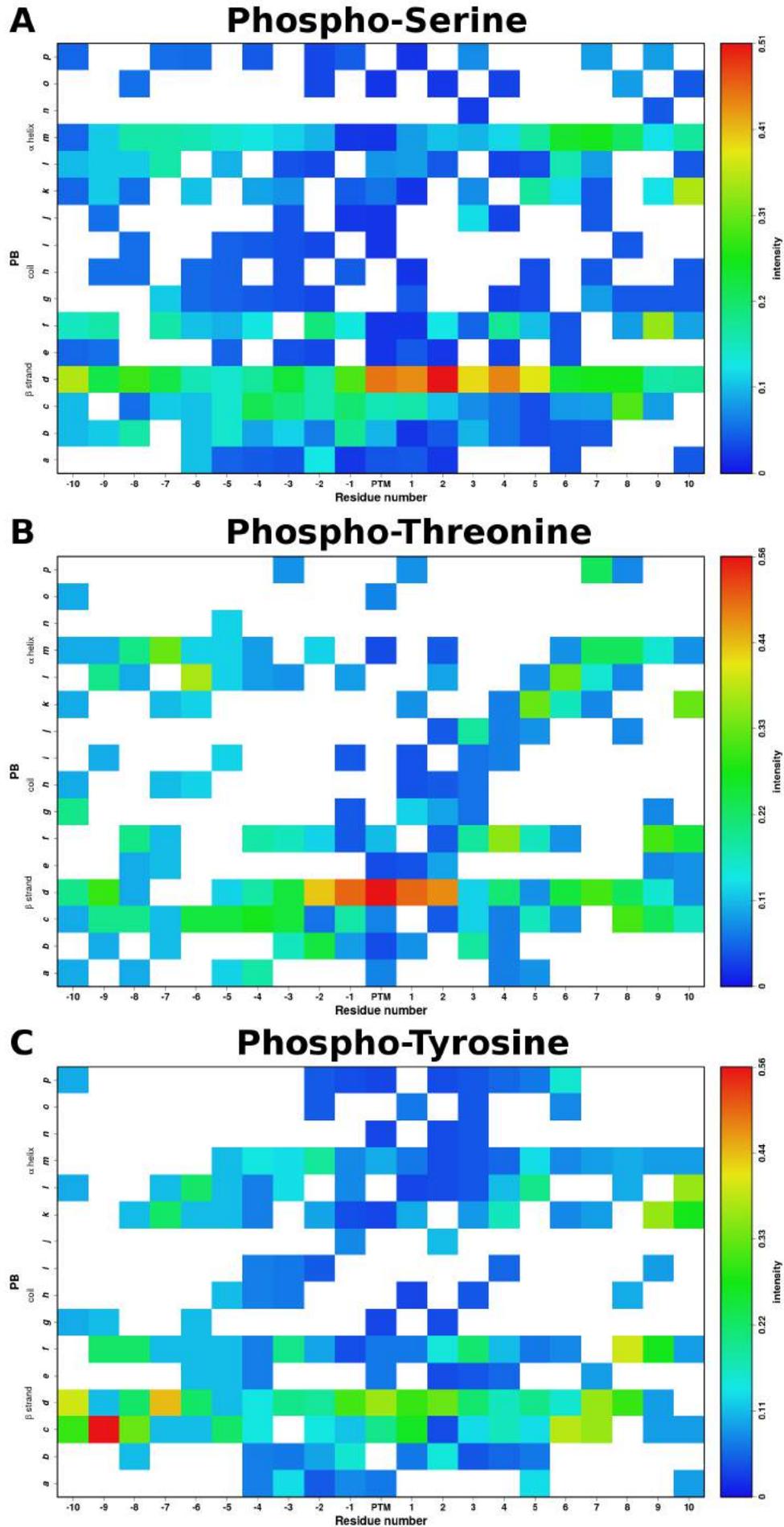

**Figure S3**. *Protein Blocks profile for phosphorylations.* The PB distributions of the complete analysis in case of **A)** Phospho-Serine, **B)** Phospho-Threonine, C) Phospho-Tyrosine. The color is represented as per the intensity legend provided on the right side. Red represents very high presence of the PB while blue being the opposite.

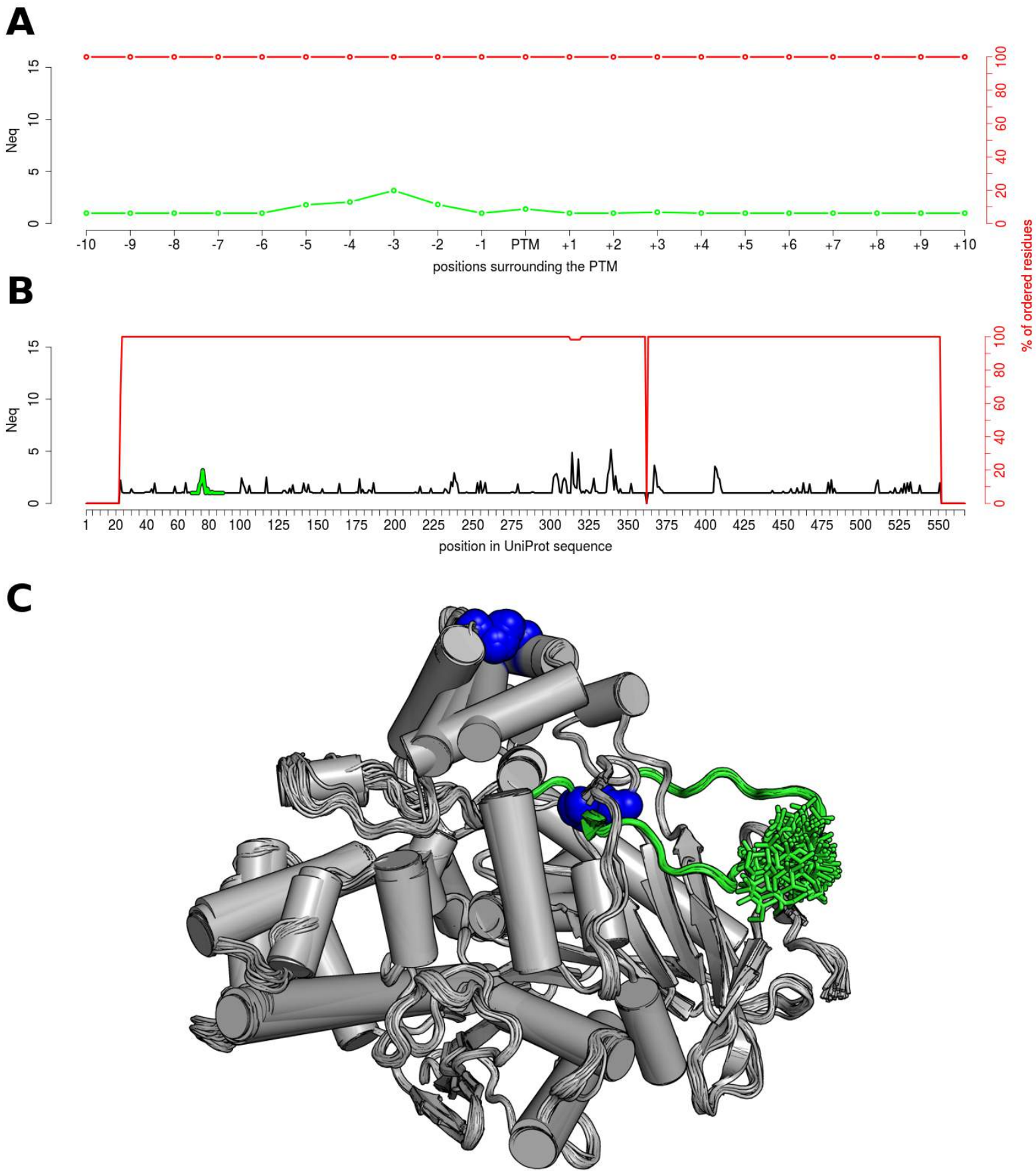

**Figure S4**. *Structural Analysis of Carboxylesterase 1*. **A)** The Neq profile of the N-linked Glycosylation at Asn 79 at its neighbouring positions. **B)** The entire Neq profile of the Carboxylesterase chain. The dip in the red curve is interpreted as the absence of data at the position. C) The structural superposition of all available crystal structures of Carboxylesterase. The green color highlights the glycosylation on the Asn 79. The blue spheres mark the disulphide bridges.

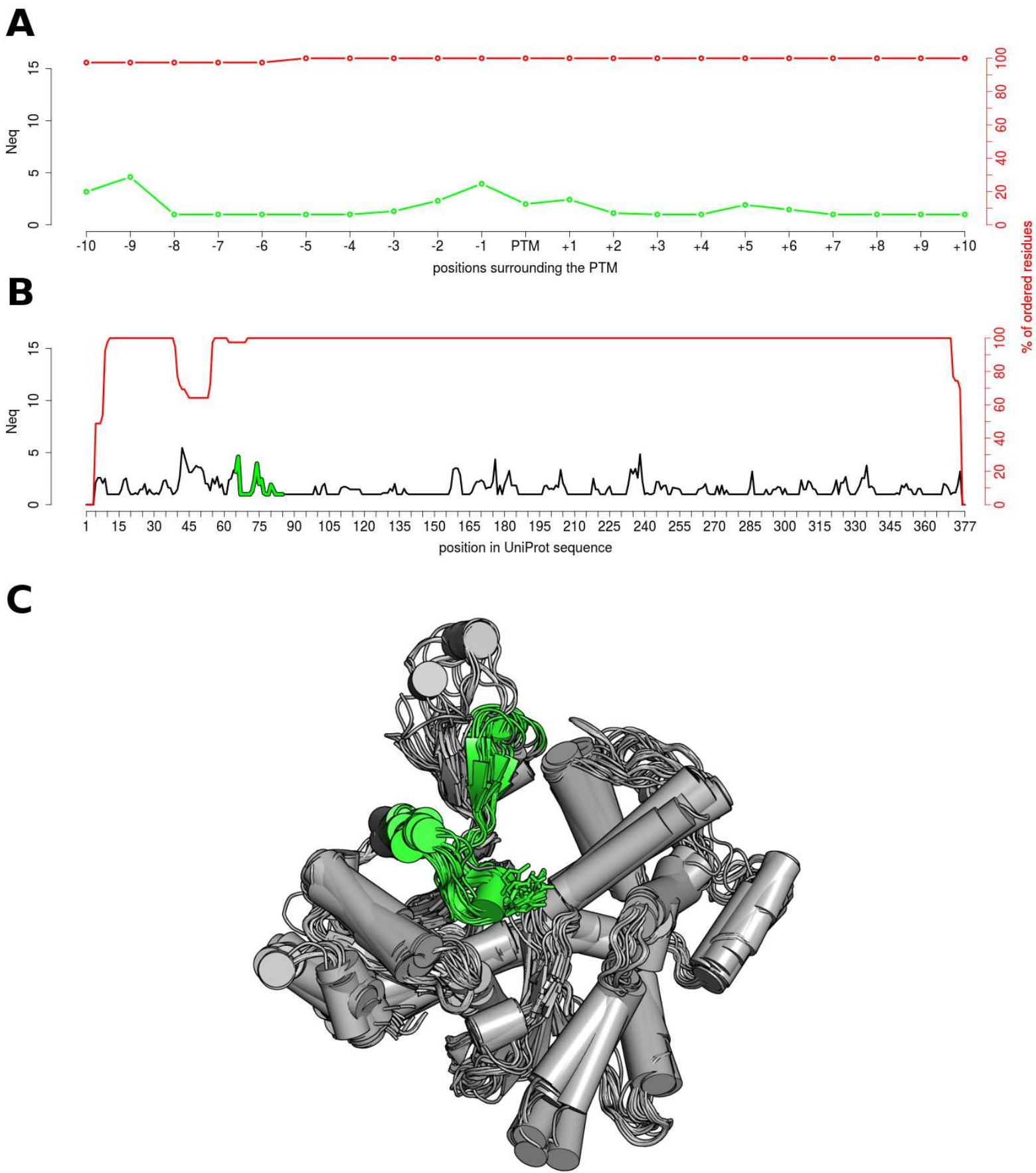

**Figure S5**. *Structural Analysis of Actin*. **A)** The $N_{eq}$ profile of the methylation of Histidine residue 75 and its neighbouring positions. **B)** The entire $N_{eq}$ profile of the actin chain. The red curve depicts the amount of crystal coordinates available for $N_{eq}$ calculation. **C)** The structural superposition of all available crystal structures (39 chains) of actin. The green color highlights the conformational change due to methylation of H75.

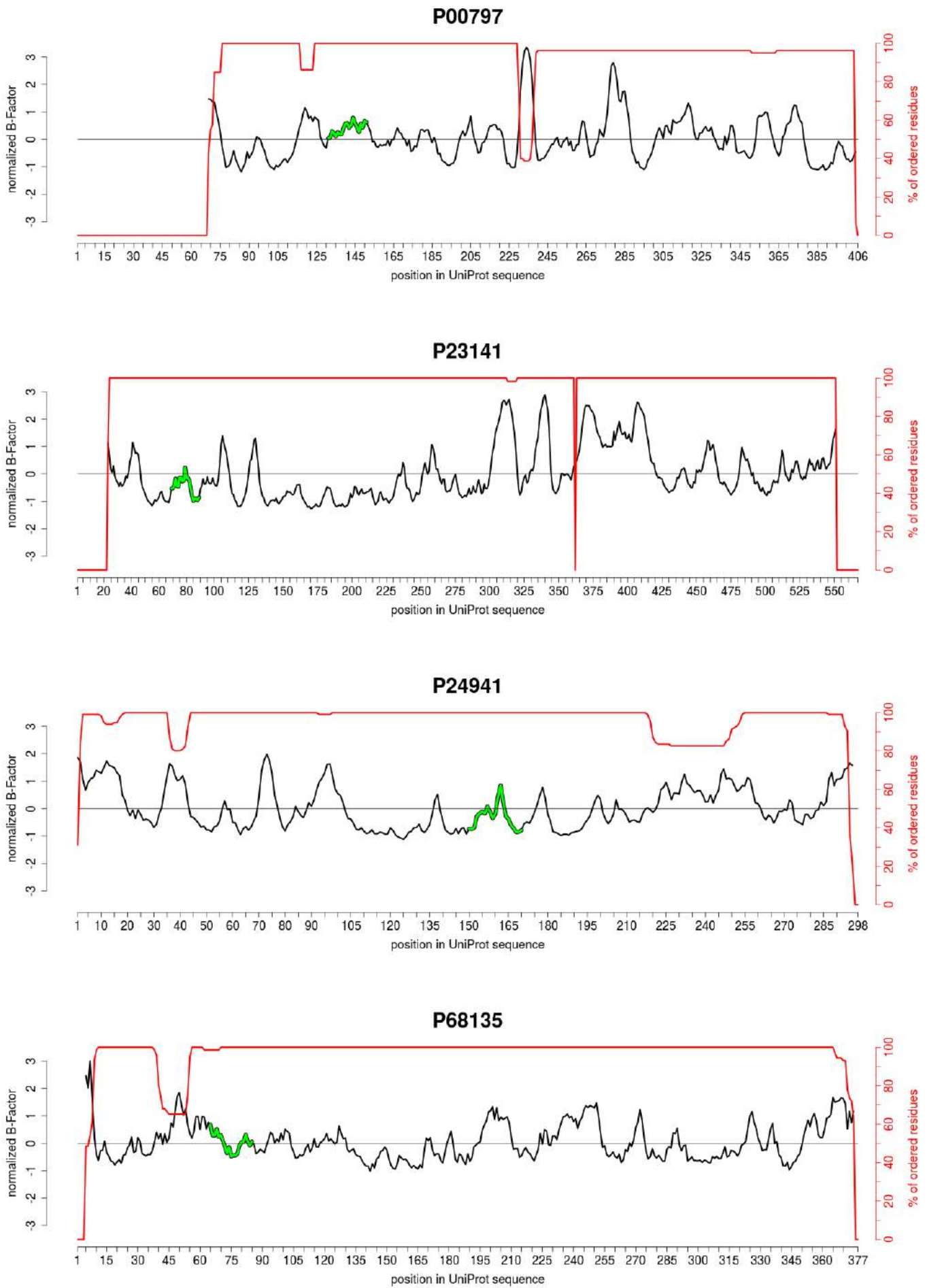

**Figure S6.** *Normalized B-Factor distribution for different PTMs*. B-factors are used to assess the extent of flexible motions present in the crystal structure. The B-factor trends match with the trends observed in $N_{eq}$ analysis. Here normalized B-factor are represented for all proteins used in the current study, containing different PTMs. **P00797** (Renin endopeptidase, N-glycosylation), **P23141** (Liver carboxylesterase 1, N-glycosylation), **P24941** (Cyclin dependent kinase 2, phosphorylation), **P68135** (Actin, methylation). The green curve highlights the PTM site and its neighbourhood.

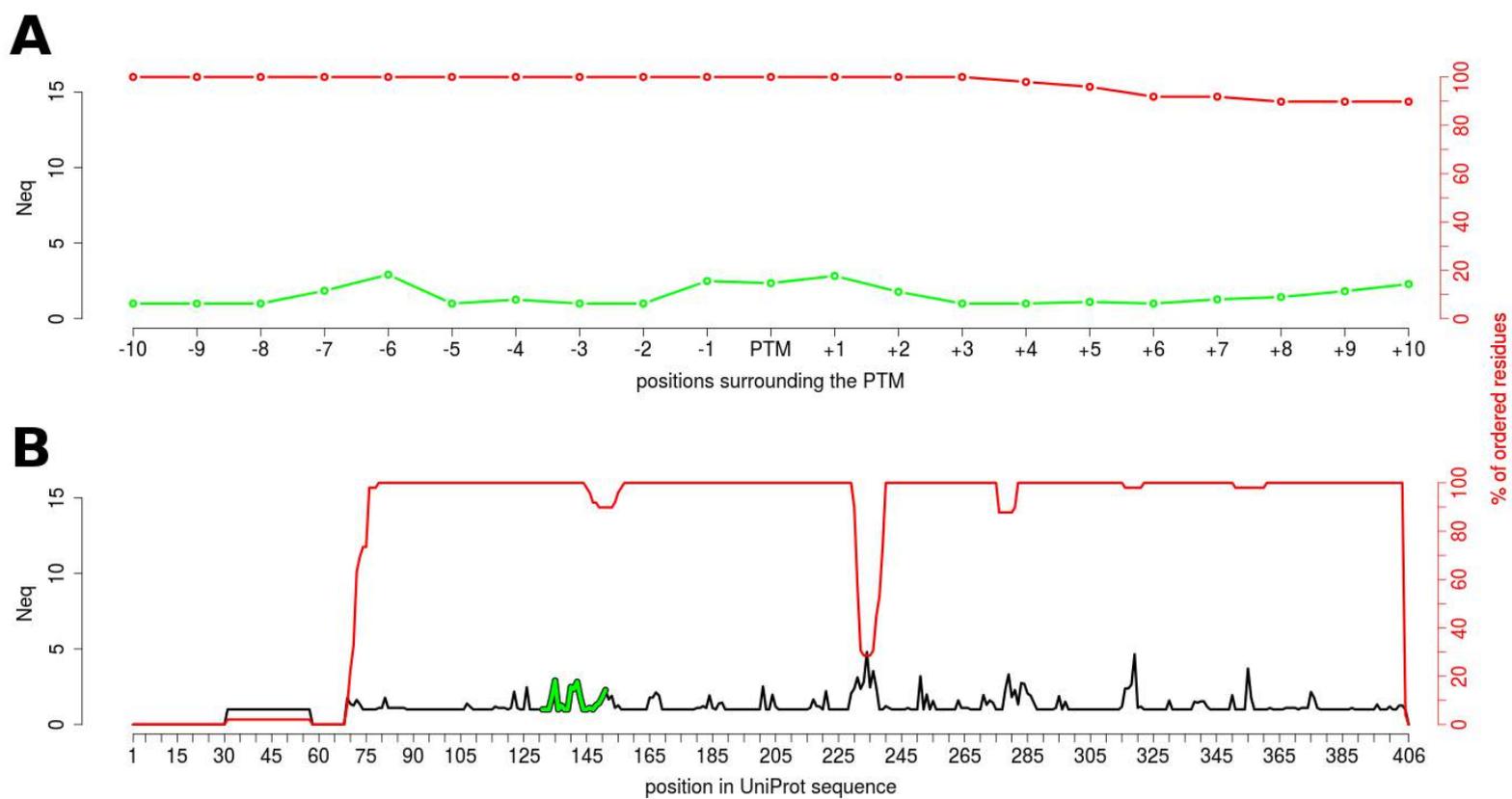

**Figure S7**. *Detailed $N_{eq}$ Profile for Renin endopeptidase without N-glycosylation at N141.* **A.** $N_{eq}$ around the PTM position (in green) with the occurrence of non-missing residue (in red). B. Same information on the complete sequence, $N_{eq}$ is in black except for the PTM position (in green, see A).

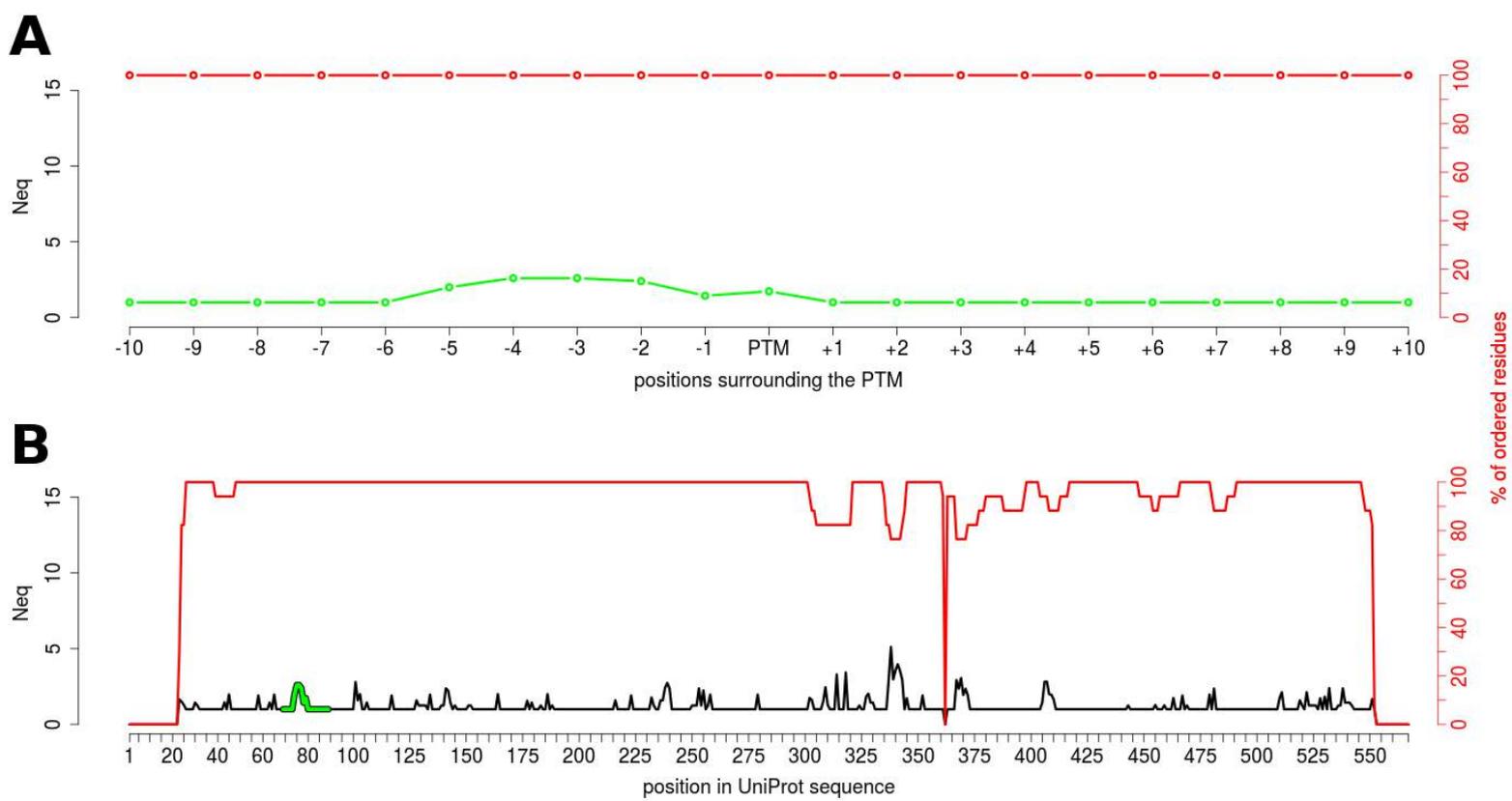

**Figure S8**. *Detailed $N_{eq}$ Profile for Liver carboxylesterase 1 without N-glycosylation at N79.* See Figure S7 for more details.

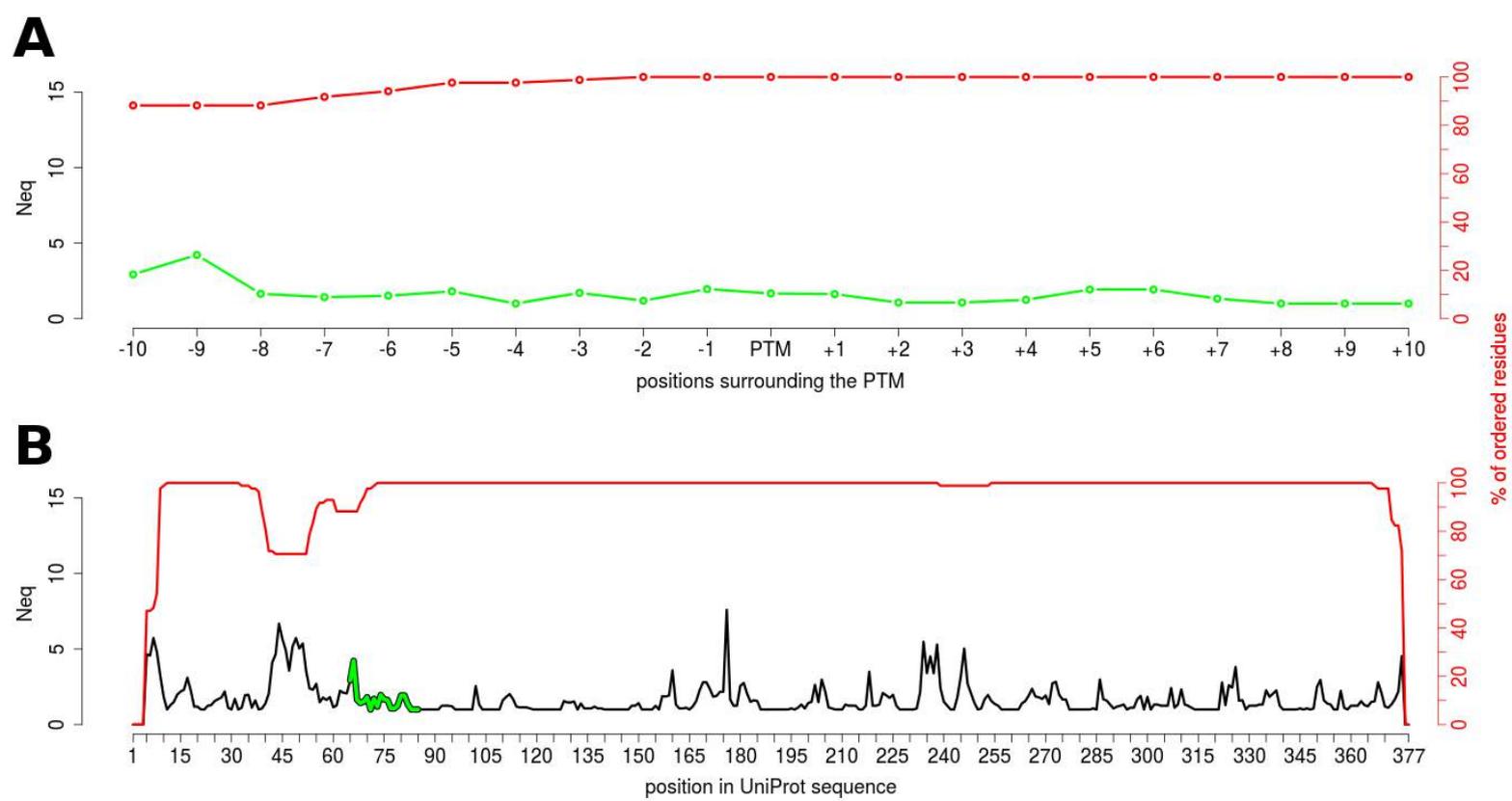

**Figure S9**. *Detailed $N_{eq}$ Profile for Actin without methylation at H75*. See Figure S7 for more details.

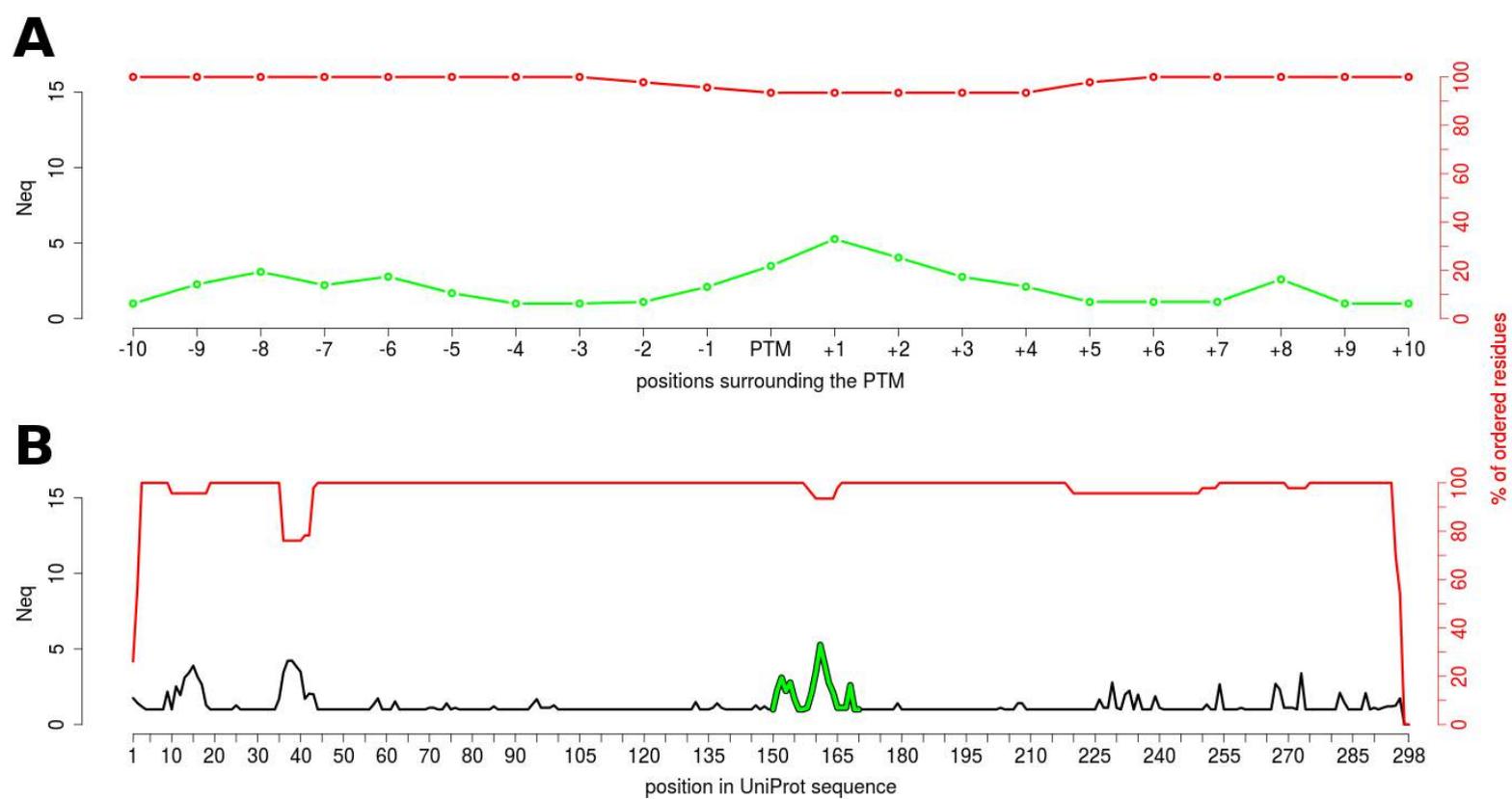

**Figure S10**. *Detailed $N_{eq}$ Profile for CDK2 without Phosphorylation at T 160.* See Figure S7 for more details.

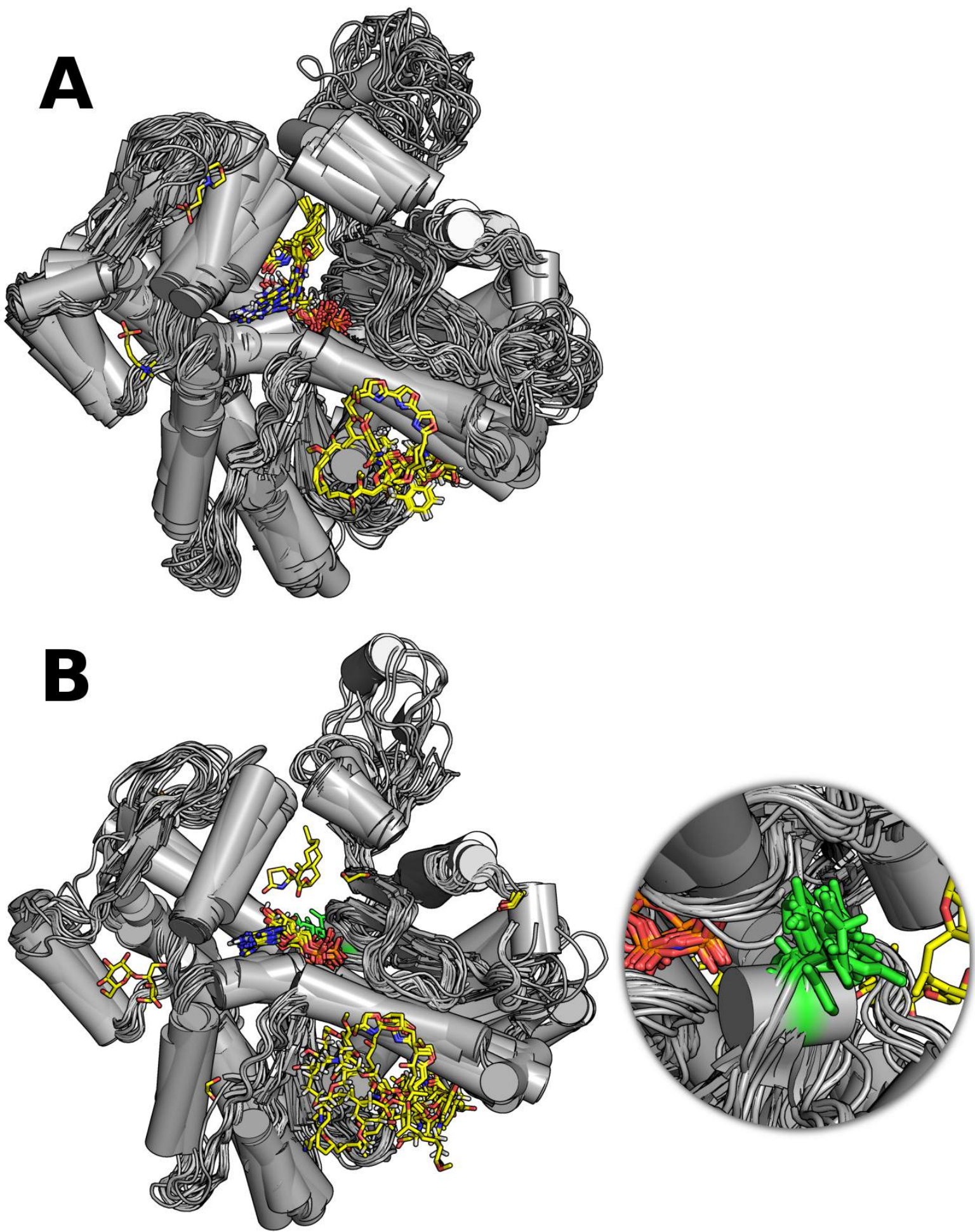

**Figure S11**. *Impact of methylation on Actin and its ligand binding.* Superimposed Actin structure along with its binding ligands. **A)** Actin with its ligand, without the methylation of H75. The ligands are represented as stick models. **B)** Actin with ligands while Histidine at 75 is methylated. Subtle changes can be observed in binding pattern of the ligand due to the presence of PTM. However, the data-set does not allow to precise the nature of such changes.

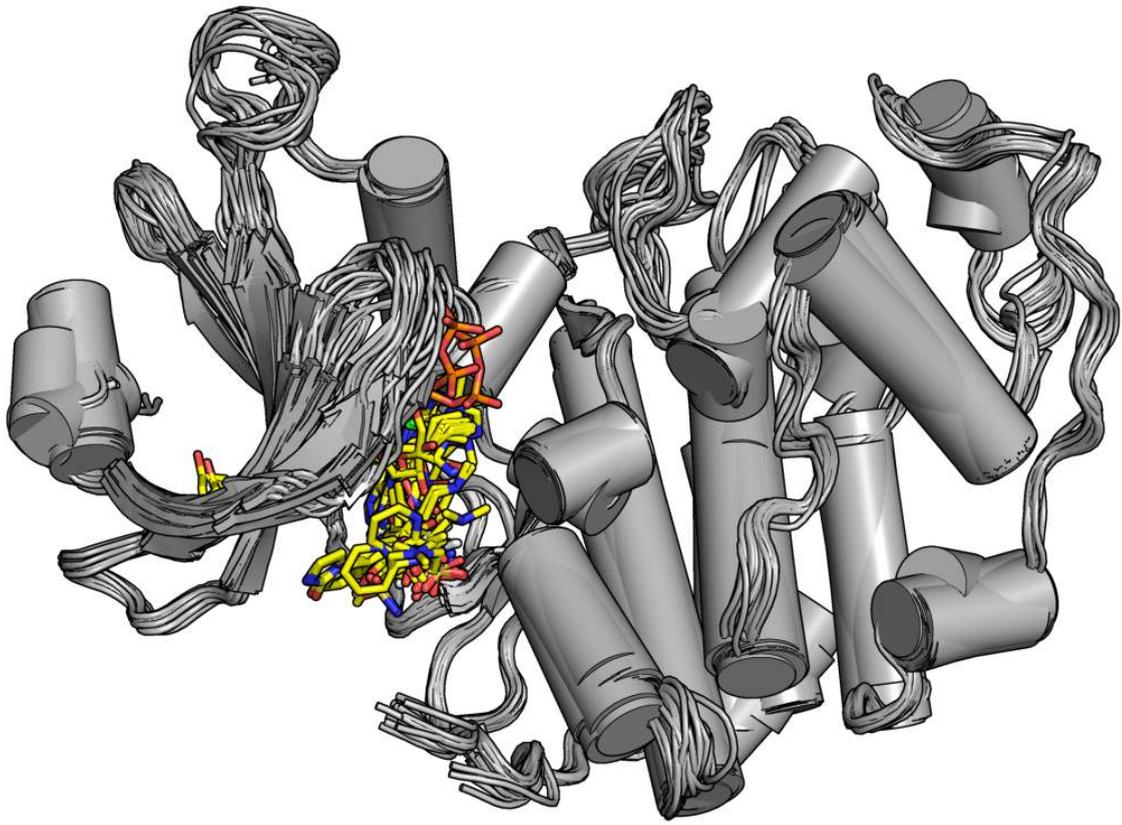

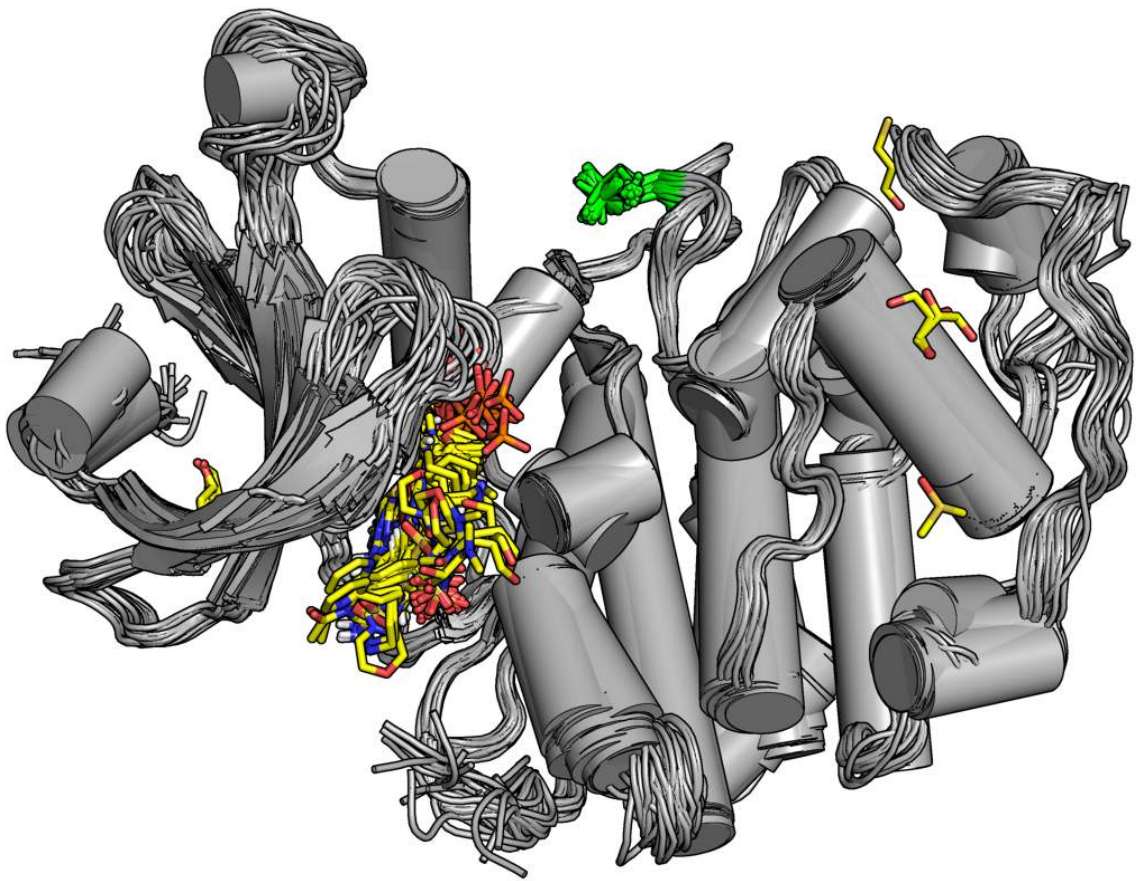

**Figure S12**. *Impact of phosphorylation on CDK on the ligand binding.* Superimposed CDK2 structure along with Cyclin A2 (regulator) bound. **A)** CDK2 with Cyclin A2, without the phosphorylation of T169. The cyclin is represented as stick model. **B)** CDK2 with Cyclin A2 while Threonine at 169 is phosphorylated. No visible change is observed in binding pattern of the ligand due to the presence or absence of the PTM.

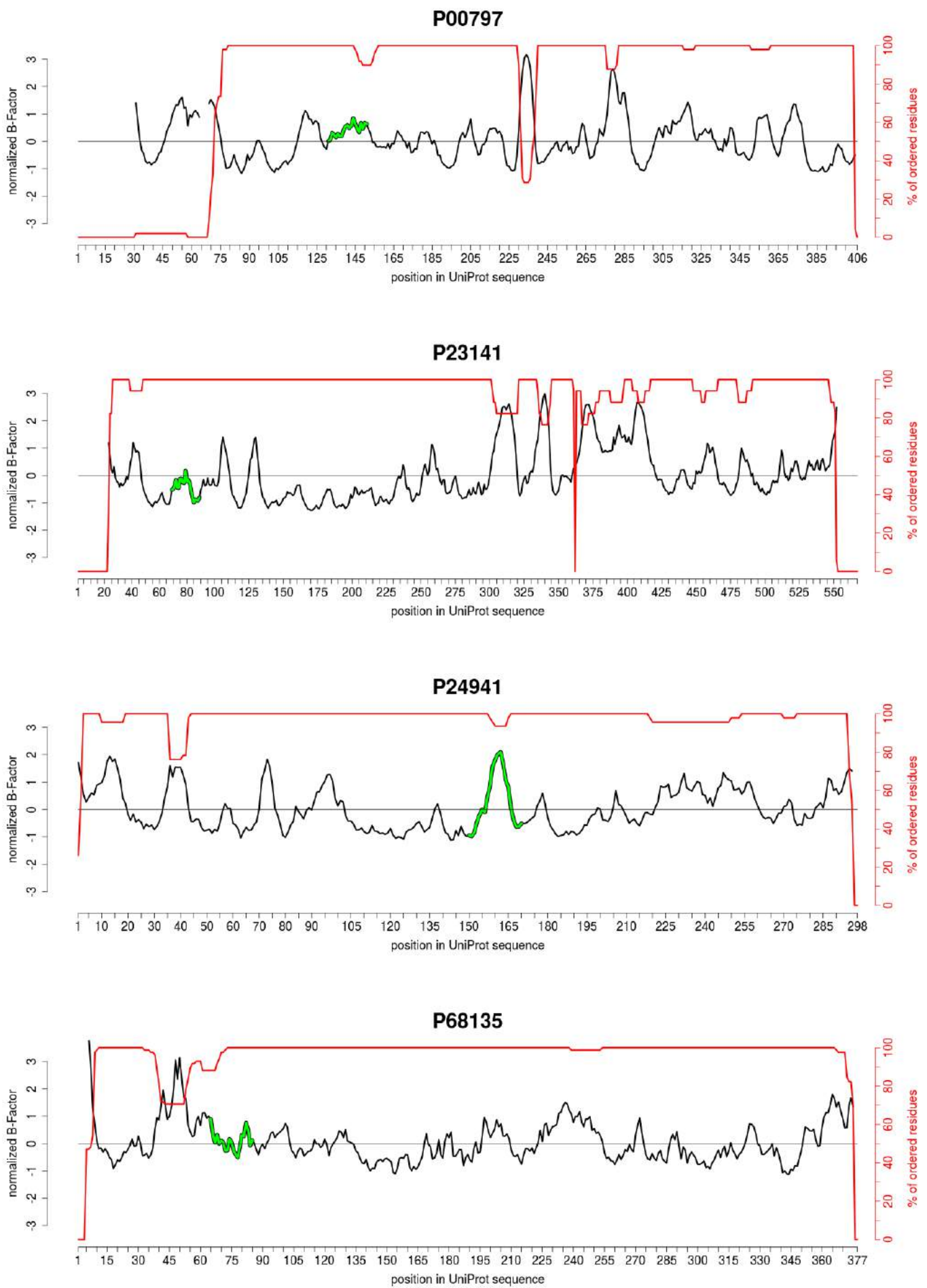

**Figure S13**. *Normalized B-Factor distribution for different proteins without PTM*. Normalized B-factor are represented for all proteins used in the current study, without any PTMs. **P00797** (Renin endopeptidase, N-glycosylation), **P23141** (Liver carboxylesterase 1, N-glycosylation), **P24941** (Cyclin dependent kinase 2, phosphorylation), **P68135** (Actin, methylation). The green curve highlights the PTM site and its neighbourhood. When compared to Figure S6, major structure impacted due to PTM is observed to be CDK2, especially in the PTM neighbourhood.

**Table S1.** *The Dataset for PTM analysis.* Using PTM-SD, a comprehensive structural dataset is prepared with PTMs, N-glycosylation, phosphorylation and methylation. The table indicates the details of the dataset with diversity indicated as no. of different source organisms, size depicted by the no. of chains and quality of data is indicated by No. of PTM. Similar statistic is also given for the derived non-redundant dataset (in columns 5 to 7).

| PTM type | Whole data | | | Non-redundant dataset | | |
|---|---|---|---|---|---|---|
| | Number of organisms | Number of chains | Number of PTMs | Number of organisms | Number of chains | Number of PTMs |
| N-glycosylation | 100 | 3092 | 7110 | 41 | 156 | 348 |
| phosphorylation | 22 | 1 308 | 1 874 | 12 | 75 | 92 |
| methylation | 21 | 584 | 886 | 9 | 15 | 19 |

**Table S2.** *The Dataset for Phosphorylation analysis:* The table represents the details of the dataset comprising of different kind of phosphorylation modifications, built using PTM-SD. The diversity of the data is indicated by the no. of different source organisms, size depicted by the no. of chains and quality of data is indicated by No. of PTM. Similar statistics is also given for the derived non-redundant dataset (in columns 5 to 7).

| PTM type | Whole data | | | Non-redundant data | | |
|---|---|---|---|---|---|---|
| | Number of organisms | Number of chains | Number of PTMs | Number of organisms | Number of chains | Number of PTMs |
| phospho-serine | 14 | 498 | 618 | 8 | 45 | 57 |
| phospho-threonine | 11 | 561 | 611 | 7 | 29 | 29 |
| phospho-tyrosine | 6 | 453 | 638 | 6 | 30 | 34 |

**Table S3.** *Dataset to analyse local and global impacts of PTMs on 4 proteins.* Four proteins as listed in first column, they are selected to study the impact of PTM on the protein structure. Second Column lists the modification taken into account while third and fourth columns are the number of structures used for comparison of structural impact in presence and absence of the PTM, respectively.

| Proteins (UniProt-AC) | PTM type and position in sequence | Number of chains with PTM | Number of chains without PTM |
|---|---|---|---|
| Renin endopeptidase P00797 (Human) | N-glycosylation on Asn 141 | 80 | 49 |
| Liver carboxylesterase 1 P23141 (Human) | N-glycosylation on Asn 79 | 59 | 17 |
| Cyclin-dependent kinase P24941 (Human) | Phosphorylation on Thr 160 | 96 | 46 |
| Actin P68135 (Rabbit) | Methylation on His 75 | 39 | 85 |

**Table S4.** *Comparison of structures with or without PTMs. Statistical tests for the 4 proteins.* Shapiro - Wilk (SW) test checks if it follows Normal law distribution, while Mann – Whitney - Wilcoxon (MWW) is a nonparametric test that compared mean values. Are indicated the size of samples (n), the calculated statistics (stats), and the p-values. The chosen risk α is equal to 5%, the significant values allow dismissing the hypothesis H0 and are colored in red.

| | | SW *Bfactor* rest of the protein vs Normal law | MWW *Bfactor* PTM region vs *Bfactor* rest of the protein | SW *Bfactor* all protein with 0 PTM vs Normal law | SW *Bfactor* all protein with 1 PTM vs Normal law | MWW *Bfactor* 0 PTM vs *Bfactor* 1 PTM | |
|---|---|---|---|---|---|---|---|
| Renin endopeptidase P00797 (Human) N-glycosylation | *n* | 21 | 316 | 21/316 | 406 | 406 | 406/406 |
| | Statistic | 0.9615 | 0.9116 | 4804 | 0.9447 | 0.9195 | 64262 |
| | *p*-value | **5.46E-01** | **1.13E-12** | **5.90E-04** | **1.51E-10** | **1.81E-12** | **5.20E-01** |
| Liver carboxylesterase 1 P23141 (Human) N-glycosylation | *n* | 21 | 507 | 21/507 | 567 | 567 | 567/567 |
| | Statistic | 0.9319 | 0.9035 | 4006 | 0.9033 | 0.8991 | 141442 |
| | *p*-value | **1.51E-01** | **2.20E-17** | **5.46E-02** | **8.34E-18** | **3.57E-18** | **7.19E-01** |
| Cyclin-dependent kinase P24941 (Human) phosphorylation | *n* | 21 | 275 | 21/275 | 298 | 298 | 298/298 |
| | Statistic | 0.9045 | 0.9383 | 2160.0000 | 0.9392 | 0.9351 | 43323 |
| | *p*-value | **4.27E-02** | **2.61E-09** | **5.45E-02** | **1.12E-09** | **4.26E-10** | **8.16E-01** |
| Actin P68135 (Rabbit) methylation | *n* | 21 | 348 | 21/348 | 377 | 377 | 377/377 |
| | Statistic | 0.9340 | 0.8969 | 3794 | 0.8915 | 0.8949 | 68532 |
| | *p*-value | **1.65E-01** | **1.32E-14** | **7.69E-01** | **1.45E-15** | **2.94E-15** | **9.78E-01** |

The experimental methods and resolutions are distributed from the PDB search result page as follow :

For the **PTM** analysis :

- • Experimental Method:
    - o X-ray (387)
    - o Electron Microscopy (8)
    - o Solution NMR (7)
    - o Fiber Diffraction (1)
- • X-ray Resolution:
    - o less than 1.5 Å (10)
    - o 1.5 - 2.0 Å (82)
    - o 2.0 - 2.5 Å (142)
    - o 2.5 - 3.0 Å (114)
    - o 3.0 and more Å (39)

For the **no-PTM** analysis :

- • Experimental Method
    - o X-ray (306)
    - o Electron Microscopy (7)
    - o Electron Crystallography (1)
- • X-ray Resolution
    - o less than 1.5 Å (4)
    - o 1.5 - 2.0 Å (153)
    - o 2.0 - 2.5 Å (105)
    - o 2.5 - 3.0 Å (37)
    - o 3.0 and more Å (7)

PDB id for PTM :

1A39
1A4G
1AC5
1AGM
1AK0
1AOT
1AQ0
1ATN
1B37
1B5F

1BHG
1BLF
1BTE
1BU8
1BVW
1BY2
1CE7
1CF3
1CX8
1CZF
1D4W
1DEO
1DMT
1DYM
1E9H
1ESV
1F42
1F88
1FJR
1FQ1
1G5G
1GH7
1GY3
1H1P
1H1Q
1H1R
1H1S
1HLG
1HNF
1I7W
1IB1
1ICF
1IKO
1IRS
1ITQ
1J4L
1J4P
1J6Z
1JMA
1JND
1JST
1JSU
1JU5
1JUH
1KSI
1KTB
1L6Z
1L8J
1LF8
1LOT

1LQV
1MA9
1MN1
1MX1
1MX5
1MYR
1N26
1N73
1NOU
1NWK
1NXC
1OT5
1OVA
1OZN
1P22
1P5E
1PC8
1PKD
1PY1
1QFX
1QG1
1QMZ
1QZ5
1QZ6
1R42
1REO
1RJ5
1RMG
1RPA
1S4N
1SCH
1T15
1T7V
1TH1
1TU5
1U5Q
1U9I
1UUR
1VSG
1W98
1WD3
1WPX
1WUA
1X27
1XWD
1Y1E
1Y64
1YA4
1YA8
1YAE

1YAH
1YAJ
1YVH
1YWN
1YXQ
1Z68
1Z6I
1Z8D
1ZAG
1ZPU
2A3Z
2A40
2A41
2A42
2AC1
2AHX
2AW2
2B5I
2BVA
2C6T
2CCH
2CCI
2D1K
2D7I
2DF3
2DQY
2DQZ
2DR0
2E4U
2E56
2ERJ
2ETZ
2FXU
2G1N
2G1O
2G1R
2G1S
2G1Y
2G22
2G9X
2GJX
2GUY
2GWJ
2GWK
2GY5
2H7C
2HCZ
2HDX
2HMH
2HMP

2HOR
2HRQ
2HRR
2I6S
2IH8
2IW6
2IW8
2IW9
2J0L
2JFL
2JGZ
2NRU
2O8Y
2OBD
2PAV
2PBD
2PCU
2PE4
2PMV
2PSX
2Q0U
2Q97
2QC1
2REN
2UZB
2UZD
2W3O
2WMA
2WMB
2Y1K
2Z64
2Z81
2ZWH
3A77
3A7F
3AL3
3B1B
3B3Q
3BGM
3BHB
3BHT
3BHU
3BHV
3BN3
3BUN
3BZI
3C4W
3CFW
3CI5
3CIG

3CL4
3COJ
3D12
3D2D
3DDP
3DDQ
3DI3
3DOG
3EXH
3F6K
3F8U
3FBX
3FEC
3FGR
3FQU
3FQX
3FXZ
3G2V
3G37
3G5C
3G6Z
3G70
3G72
3GW5
3H5C
3HBT
3HKL
3HN3
3HUF
3INB
3IU3
3J8A
3J8F
3JVF
3JYH
3K1W
3K2L
3K4P
3K4Q
3KM4
3KQ4
3LB6
3LGD
3LKJ
3LMY
3LW1
3M1F
3M3N
3MAZ
3MDJ

3MFP
3MHR
3MJ7
3ML4
3MXC
3MY5
3NKM
3O4O
3O9L
3OAD
3OAG
3OB1
3OJY
3OL2
3OMH
3OOT
3OQF
3OQK
3Q3T
3Q4A
3Q4B
3Q5H
3QD2
3QHR
3QHW
3QS7
3S98
3SFC
3SI1
3TMP
3TNW
3TPE
3TRQ
3U7B
3UAL
3UEO
3UNN
3UZD
3V7D
3VA4
3VCM
3VSW
3VSX
3VTA
3VUC
3VYD
3VYE
3VYF
3W81
3WP1

3ZEW
3ZKF
3ZNI
4A7F
4A7H
4A7L
4A7N
4BCK
4BCM
4BCN
4BCO
4BCP
4BCQ
4C4E
4CFM
4CFN
4CFU
4CFV
4CFW
4CFX
4CXA
4DC2
4EOI
4EOJ
4EOK
4EOL
4EOM
4EON
4EOO
4EOP
4EOQ
4EOR
4EOS
4EUU
4F7B
4FDI
4FGU
4FXW
4GDX
4GJ5
4GJ6
4GJ7
4GJ8
4GJ9
4GJA
4GJB
4GJC
4GJD
4GL9
4GLR

4GVC
4GZ9
4H03
4H1S
4I3Z
4IAN
4IGK
4II5
4J6S
4JAX
4JLU
4JMG
4JMH
4JS1
4JS8
4KIK
4KJY
4L1U
4MHX
4NM5
4NM7
4NO3
4NU1
4OTD
4PO7
4PSI
4PYV
4Q1N
4QSY
4R3P
4R3R
4R4H
4RA0
4RER
4RH5
4RXZ
4RYC
4RYG
4RZ1
4S1G
4V11
5A7F
5A7G

PDB id for non-PTM :

1AQ1
1B38
1B39
1BBS
1BIL
1BIM
1CKP
1DI8
1DM2
1E1V
1E1X
1FIN
1FVT
1FVV
1G5S
1GIH
1GII
1GIJ
1GZ8
1H00
1H01
1H07
1H08
1H0V
1H0W
1HCK
1HCL
1IJJ
1JVP
1KE5
1KE6
1KE7
1KE8
1KE9
1LCU
1MX9
1OGU
1OI9
1OIQ
1OIR
1OIT
1OIU
1OIY
1P2A
1PF8
1PW2
1PXI
1PXJ

1PXK
1PXL
1PXM
1PXN
1PXO
1PXP
1PYE
1R78
1RDW
1RFQ
1URW
1V1K
1VYW
1VYZ
1W0X
1W8C
1WCC
1Y8Y
1Y91
1YKR
2A0C
2A4L
2A5X
2B52
2B54
2B55
2BHE
2BHH
2BKZ
2BPM
2BTR
2BTS
2C4G
2C5X
2C5Y
2C68
2C69
2C6I
2C6K
2C6L
2C6M
2C6O
2CLX
2DS1
2DUV
2EXM
2FS4
2FVD
2G20
2G21

2G24
2G26
2G27
2I40
2I4Q
2IKO
2IKU
2IL2
2J9M
2Q1N
2Q31
2Q36
2R64
2UZN
2UZO
2V0D
2VTA
2VTH
2VTI
2VTJ
2VTL
2VTM
2VTN
2VTO
2VTP
2VTQ
2VTR
2VTS
2VTT
2VU3
2VV9
2W05
2W06
2W17
2W1H
2WIH
2WIP
2WPA
2WXV
2XMY
2XNB
2Y83
3B5U
3D91
3EID
3EJ1
3EOC
3EZR
3EZV
3F5X

3FZ1
3IG7
3IGG
3J4K
3J8I
3J8J
3J8K
3JBJ
3K9B
3LE6
3LFN
3LFQ
3LFS
3M6G
3NS9
3OWN
3PJ8
3PXF
3PXQ
3PXR
3PXY
3PXZ
3PY0
3PY1
3QL8
3QQF
3QQG
3QQH
3QQJ
3QQK
3QQL
3QRT
3QRU
3QTQ
3QTR
3QTS
3QTU
3QTW
3QTX
3QTZ
3QU0
3QWJ
3QWK
3QX2
3QX4
3QXO
3QXP
3QZF
3QZG
3QZH

3QZI
3R1Q
3R1S
3R1Y
3R28
3R6X
3R71
3R73
3R7E
3R7I
3R7U
3R7V
3R7Y
3R83
3R8L
3R8M
3R8P
3R8U
3R8V
3R8Z
3R9D
3R9H
3R9N
3R9O
3RAH
3RAI
3RAK
3RAL
3RJC
3RK5
3RK7
3RK9
3RKB
3RM6
3RM7
3RMF
3RNI
3ROY
3RPO
3RPR
3RPV
3RPY
3RZB
3S00
3S0O
3S1H
3S2P
3SQQ
3SW7
3TI1

3TIY
3TIZ
3TPQ
3ULI
3UNJ
3UNK
3WBL
4AB1
4ACM
4AMT
4BGH
4BZD
4D1X
4D1Z
4EK5
4EK6
4ERW
4EZ3
4EZ7
4FKG
4FKI
4FKJ
4FKO
4FKQ
4FKR
4FKS
4FKT
4FKU
4FKV
4FKW
4FX3
4GCJ
4K41
4K42
4K43
4KD1
4LYN
4RJ3
4XX3
4XX4
5A14
5A7H
5AND
5ANE
5ANG
5ANI
5ANJ
5ANK
5ANO
5CYI

5D1J
5FP5
5FP6
5IEV
5IEX
5IEY
5IF1
5JLF
5K4J
5KOQ
5KOS
5SXN
5SY2
5SY3
5SZ9
5T4S